\documentclass[aps,pra,floatfix,showpacs,preprint,onecolumn]{revtex4-2}
\usepackage{graphicx}
\usepackage{amsmath, amsfonts, amssymb, bm}

\usepackage{amstext}
\usepackage{mathrsfs}
\usepackage{hyperref}
\usepackage{cleveref}

\usepackage{color}

\begin{document}
\title{Analytical spectrum of nonlinear Thomson scattering including radiation reaction}
\author{A.~Di Piazza}
\email{dipiazza@mpi-hd.mpg.de}
\author{G. Audagnotto}
\affiliation{Max Planck Institute for Nuclear Physics, Saupfercheckweg 1, D-69117 Heidelberg, Germany}

\begin{abstract}
Accelerated charges emit electromagnetic radiation and the consequent energy-momentum loss alters their trajectory. This phenomenon is known as radiation reaction and the Landau-Lifshitz (LL) equation is the classical equation of motion of the electron, which takes into account radiation-reaction effects in the electron trajectory. By using the analytical solution of the LL equation in an arbitrary plane wave, we compute the analytical expression of the classical emission spectrum via nonlinear Thomson scattering including radiation-reaction effects. Both the angularly-resolved and the angularly-integrated spectra are reported, which are valid in an arbitrary plane wave. Also, we have obtained a phase-dependent expression of the electron dressed mass, which includes radiation-reaction effects. Finally, the corresponding spectra within the locally constant field approximation have been derived.
\end{abstract}

\maketitle

\section{Introduction}
Maxwell's and Lorentz equations allow one in principle to describe self-consistently the classical dynamics of electric charges and their electromagnetic field. However, even in the case of a single elementary charge, an electron for definiteness, the solution of the self-consistent problem of the electron dynamics and of that of its own electromagnetic field is plagued by physical inconsistencies, which ultimately are related to the divergent self energy of a point-like charge. In fact, the inclusion of the ``reaction'' of the self electromagnetic field on the electron dynamics (known as radiation reaction) implies an unavoidable Coulomb-like divergence when one evaluates the self field at the electron position \cite{Jackson_b_1975,Landau_b_2_1975,Barut_b_1980,Rohrlich_b_2007}. However, this divergence can be reabsorbed via a redefinition of the electron mass, which ultimately leads to one of the most controversial equations in physics, the Lorentz-Abraham-Dirac (LAD) equation \cite{Abraham_b_1905,Lorentz_b_1909,Dirac_1938}. In the case of interest here, where the external force is also electromagnetic, the LAD equation can be derived by eliminating from the Maxwell-Lorentz system of equations the electromagnetic field generated by the electron. In this respect, solving the LAD equation amounts to solving exactly the electron dynamics in the external electromagnetic field and plugging the resulting solution into the Li\'{e}nard-Wiechert potentials amounts to determining the corresponding exact electromagnetic field.

Even after the absorption of the divergent electron self-energy via the classical mass renormalization, the LAD equation remains problematic as it allows for so-called runaway solutions, where the electron's acceleration may exponentially increase with time even if the external field, for example, vanishes identically \cite{Jackson_b_1975,Landau_b_2_1975,Barut_b_1980,Rohrlich_b_2007}. The origin of the existence of the runaway solutions is precisely a term in the radiation-reaction force, known as the Schott term, which depends on the time-derivative of the electron acceleration, thus rendering the LAD equation a third-order differential equation with non-Newtonian features. 

Landau and Lifshitz realized that within the realm of classical electrodynamics, i.e., if quantum effects are negligible, the radiation-reaction force in the instantaneous rest frame of the electron is always much weaker than the Lorentz force \cite{Landau_b_2_1975}. This allows one to replace the electron four-acceleration in the radiation-reaction four-force with its leading-order expression, i.e., with the Lorentz four-force divided by the electron mass \cite{Landau_b_2_1975}. It is important to stress that the ``reduction of order'' proposed by Landau and Lifshitz is such that neglected quantities are much smaller than corrections induced by quantum effects, which are already ignored classically. The resulting equation is known as the Landau-Lifshitz (LL) equation and it is free of the physical inconsistencies of the LAD equation \cite{Spohn_2000}. The equivalence between the LL equation and the LAD equation within the realm of classical electrodynamics has to be intended as these equations differ by terms much smaller than quantum corrections (see Refs. \cite{Koga_2004,Bulanov_2011} for numerical tests about this equivalence and Ref. \cite{Hadad_2010} for a numerical example where the predictions of the LL and the LAD equations differ but where quantum effects indeed are large). Presently the LL equation, as well as the problem of radiation reaction in general, is being investigated by several groups both theoretically \cite{Vranic_2014,Blackburn_2014,Tamburini_2014,Li_2014,Heinzl_2015,Yoffe_2015,Capdessus_2015,Vranic_2016,
Dinu_2016,Di_Piazza_2017,Harvey_2017,Ridgers_2017,Niel_2018a,Niel_2018b} and experimentally \cite{Wistisen_2018,Cole_2018,Poder_2018} (see also the recent reviews \cite{Hammond_2010,Di_Piazza_2012,Burton_2014,Blackburn_2020} for previous publications).

Here, we present analytical expressions of the energy emission spectrum of an electron driven by an external intense plane wave (nonlinear Thomson scattering) by taking into account radiation-reaction effects via the LL equation. To achieve this goal, we use the analytical solution of the LL equation in an arbitrary plane wave \cite{Di_Piazza_2008_a} and we derive the angularly-resolved and the angularly-integrated energy spectra as double integrals over the phase of the plane wave. In this respect, we point out that additional classical corrections to the energy spectra, which would be brought about, i.e., by using the LAD equation would be smaller than already ignored quantum corrections. Finally, the corresponding expressions within the so-called locally constant field approximation (LCFA) are derived as single phase-integrals \cite{Ritus_1985,Baier_b_1998,Di_Piazza_2012}. These results obtained here also complement the ones obtained in Ref. \cite{Di_Piazza_2018_b}, where the analytical expression of the infrared limit of the emission spectrum including radiation-reaction effects was presented.

Units with $\hbar=c=4\pi\epsilon_0=1$ are employed throughout and the metric tensor is $\eta^{\mu\nu}=\text{diag}(+1,-1,-1,-1)$.

\section{Analytical spectrum of nonlinear Thomson scattering}

Let us consider an electron (charge $e<0$ and mass $m$, respectively), whose trajectory is characterized by the instantaneous position $\bm{x}(t)$ and the instantaneous velocity $\bm{\beta}(t)=d\bm{x}(t)/dt$. The electromagnetic energy $\mathcal{E}$ radiated by the electron per unit of angular frequency $\omega$ and along the direction $\bm{n}=(\sin\vartheta\cos\varphi,\sin\vartheta\sin\varphi,\cos\vartheta)$ within a solid angle $d\Omega=\sin\vartheta d\vartheta d\varphi$ is given by [see, e.g., Eq. (14.67) in Ref. \cite{Jackson_b_1975}]
\begin{equation}
\label{dE_dodO}
\frac{d\mathcal{E}}{d\omega d\Omega}=\frac{e^2\omega^2}{4\pi^2}\left|\int_{-\infty}^{\infty} dt\,\bm{n}\times(\bm{n}\times\bm{\beta}(t))e^{i\omega(t-\bm{n}\cdot\bm{x}(t))}\right|^2,
\end{equation}
and we stress that this expression of the emitted energy is valid for an arbitrary trajectory of the electron.

Now, we assume that the electron moves in the presence of a plane-wave background field, described by the four-vector potential $A^{\mu}(\phi)=(A^0(\phi),\bm{A}(\phi))$, where $\phi=(n_0x)=t-\bm{n}_0\cdot\bm{x}$, with $n_0^{\mu}=(1,\bm{n}_0)$ and the unit vector $\bm{n}_0$ identifying the propagation direction of the plane wave itself. We decide to work in the Lorenz gauge $\partial_{\mu}A^{\mu}(\phi)=(n_0A'(\phi))=0$ with the additional condition $A^0(\phi)=0$. Here and below, the prime indicates the derivative with respect to the argument of a function. By assuming that $\lim_{\phi\to\pm\infty}\bm{A}(\phi)=\bm{0}$, then the Lorenz-gauge condition implies $\bm{n}_0\cdot\bm{A}(\phi)=0$. Thus, the four-vector potential $A^{\mu}(\phi)$ can be written as $A^{\mu}(\phi)=\sum_{j=1}^2a_j^{\mu}\psi_j(\phi)$, where the four-vectors $a_j^{\mu}$ have the form $a_j^{\mu}=(0,\bm{a}_j)$ and fulfill the orthogonality conditions $(a_ja_{j'})=-\delta_{jj'}$, with $j,j'=1,2$, and $(n_0a_j)=-\bm{n}_0\cdot\bm{a}_j=0$, and where the functions $\psi_j(\phi)$ are arbitrary (physically well-behaved) functions such that $\lim_{\phi\to\pm\infty}\psi_j(\phi)=0$. 

It is convenient first to express the emitted energy $d\mathcal{E}/d\omega d\Omega$ as an integral over the laser phase $\varphi=\omega_0\phi$, where $\omega_0$ is the central angular frequency of the plane wave (or, more in general, an arbitrary frequency scale describing the time dependence of the plane wave). This is easily done because $d\phi(t)/dt=1-\bm{n}_0\cdot\bm{\beta}(t)$ along the electron trajectory and one obtains
\begin{equation}
\label{dE_dodO_phase}
\frac{d\mathcal{E}}{d\omega d\Omega}=\frac{e^2}{4\pi^2}\frac{\omega^2}{\omega_0^2}\left|\int_{-\infty}^{\infty} d\varphi\,\frac{\bm{n}\times(\bm{n}\times\bm{p}(\varphi))}{p_-(\varphi)}e^{i\frac{\omega}{\omega_0}\int_{-\infty}^{\varphi}d\varphi'\frac{\varepsilon(\varphi')-\bm{n}\cdot\bm{p}(\varphi')}{p_-(\varphi')}}\right|^2,
\end{equation}
where $p^{\mu}(\varphi)=(\varepsilon(\varphi),\bm{p}(\varphi))=\varepsilon(\varphi)(1,\bm{\beta}(\varphi))$, with $\varepsilon(\varphi)=m/\sqrt{1-\bm{\beta}^2(\varphi)}$, is the electron four-momentum and $p_-(\varphi)=(n_0p(\varphi))$. In fact, it is in general convenient to introduce also the four-dimensional quantity $\tilde{n}_0^{\mu}=(1,-\bm{n}_0)/2$ because $n_0^{\mu}$, $\tilde{n}_0^{\mu}$, and $a^{\mu}_j$ fulfill the completeness relation: $\eta^{\mu\nu}=n_0^{\mu}\tilde{n}_0^{\nu}+\tilde{n}_0^{\mu}n_0^{\nu}-a_1^{\mu}a_1^{\nu}-a_2^{\mu}a_2^{\nu}$ (note that $(n_0\tilde{n}_0)=1$ and that, as we have already seen, $(a_1a_1)=(a_2a_2)=-1$, whereas all other possible scalar products among $n_0^{\mu}$, $\tilde{n}_0^{\mu}$, and $a^{\mu}_j$ vanish). By using the quantities $n_0^{\mu}$, $\tilde{n}_0^{\mu}$, and $a^{\mu}_j$ one can define the light-cone coordinates of an arbitrary four-vector $v^{\mu}=(v^0,\bm{v})$ as $v_+=(\tilde{n}_0v)$, $\bm{v}_{\perp}=-((va_1),(va_2))$, and $v_-=(n_0v)$. Also, the four-dimensional scalar product between two four-vectors $a^{\mu}$ and $b^{\mu}$ can be written as $(ab)=a_+b_-+a_-b_+-\bm{a}_{\perp}\cdot\bm{b}_{\perp}$.

Now, we recall that the LL equation in an external electromagnetic field $F^{\mu\nu}=F^{\mu\nu}(x)$ reads \cite{Landau_b_2_1975}
\begin{equation}
\label{iLL_eq}
\begin{split}
m\frac{d u^{\mu}}{ds}=&eF^{\mu\nu}u_{\nu}+\frac{2}{3}e^2\left[\frac{e}{m}(\partial_{\alpha}F^{\mu\nu})u^{\alpha}u_{\nu}\right.\\
&\left.+\frac{e^2}{m^2}F^{\mu\nu}F_{\nu\alpha}u^{\alpha}+\frac{e^2}{m^2}(F^{\alpha\nu}u_{\nu})(F_{\alpha\lambda}u^{\lambda})u^{\mu}\right],
\end{split}
\end{equation}
where $s$ is the electron proper time and $u^{\mu}(s)=p^{\mu}(s)/m$ is the electron four-velocity. In the case of the plane wave described above, we can introduce the central laser four-wave-vector as $k_0^{\mu}=\omega_0n_0^{\mu}$ such that the laser phase reads $\varphi=(k_0x)$. By indicating as $p_0^{\mu}=(\varepsilon_0,\bm{p}_0)$, with $\varepsilon_0=\sqrt{m^2+\bm{p}_0^2}$, the initial four-momentum of the electron, i.e., $\lim_{\varphi\to-\infty}p^{\mu}(\varphi)=p_0^{\mu}$, the four-momentum $p^{\mu}(\varphi)$ at the generic phase $\varphi$ is given by \cite{Di_Piazza_2008_a}
\begin{equation}
\label{iSol_A}
\begin{split}
p^{\mu}(\varphi)=&\frac{1}{h(\varphi)}\left\{p_0^{\mu}+\frac{1}{2\eta_0}[h^2(\varphi)-1]k_0^{\mu}+\frac{\omega_0}{m\eta_0}\mathcal{P}^{\mu}(\varphi)-\frac{\omega_0^2}{2m^4\eta^3_0}\mathcal{P}^2(\varphi)k_0^{\mu}\right\}.
\end{split}
\end{equation}
In this expression we have introduced the parameter $\eta_0=(k_0p_0)/m^2$ and the functions
\begin{align}
\label{h}
h(\varphi)&=1+\frac{2}{3}e^2\eta_0\int_{-\infty}^{\varphi}d\tilde{\varphi}\,\bm{\xi}_{\perp}^2(\tilde{\varphi}),\\
\mathcal{P}^{\mu}(\varphi)&=\mathcal{F}^{\mu\nu}(\varphi)p_{0,\nu},
\end{align}
where
\begin{equation}
\label{F_munu}
\mathcal{F}^{\mu\nu}(\varphi)=\int_{-\infty}^{\varphi}d\tilde{\varphi}\,\left[h(\tilde{\varphi})\xi^{\mu\nu}(\tilde{\varphi})+\frac{2}{3}e^2\eta_0\xi^{\prime\,\mu\nu}(\tilde{\varphi})\right],
\end{equation}
with $\bm{\xi}_{\perp}(\varphi)=(e/m)\bm{A}'_{\perp}(\varphi)$ and $\xi^{\mu\nu}(\varphi)=(e/m)[n_0^{\mu}A^{\prime\,\nu}(\varphi)-n_0^{\nu}A^{\prime\,\mu}(\varphi)]$. Note that, assuming that $|\xi^{\mu\nu}(\varphi)|\sim |\xi^{\prime\,\mu\nu}(\varphi)|$ as it is typically the case for standard laser fields, the term proportional to $\xi^{\prime\,\mu\nu}(\varphi)$ in $\mathcal{F}^{\mu\nu}(\varphi)$ can be neglected according to Landau and Lifshitz reduction of order \cite{Landau_b_2_1975} (see Ref. \cite{Khokonov_2019} for a recent study on this term and Ref. \cite{Di_Piazza_2017} for a situation where it cannot be ignored). For this reason we write
\begin{equation}
\label{F}
\mathcal{F}^{\mu\nu}(\varphi)=\int_{-\infty}^{\varphi}d\tilde{\varphi}\,h(\tilde{\varphi})\xi^{\mu\nu}(\tilde{\varphi})
\end{equation}
and we use this expression below. For the sake of later convenience, we also report here the light-cone components of the four-momentum of the electron in the plane wave including radiation reaction:
\begin{align}
\label{p_-}
p_-(\varphi)&=\frac{p_{0,-}}{h(\varphi)},\\
\bm{p}_{\perp}(\varphi)&=\frac{1}{h(\varphi)}[\bm{p}_{0,\perp}-m\bm{\mathcal{F}}_{\perp}(\varphi)],\\
\label{p_+}
p_+(\varphi)&=\frac{m^2+\bm{p}^2_{\perp}(\varphi)}{2p_-(\varphi)}=\frac{1}{h(\varphi)}\frac{m^2h^2(\varphi)+[\bm{p}_{0,\perp}-m\bm{\mathcal{F}}_{\perp}(\varphi)]^2}{2p_{0,-}},
\end{align}
where $\bm{\mathcal{F}}_{\perp}(\varphi)=\int_{-\infty}^{\varphi}d\tilde{\varphi}\,h(\tilde{\varphi})\bm{\xi}_{\perp}(\tilde{\varphi})$ [see Eq. (\ref{F})] as well as the corresponding longitudinal momentum [$p_{\parallel}(\varphi)=\bm{n}_0\cdot\bm{p}(\varphi)$] and the energy:
\begin{align}
p_{\parallel}(\varphi)&=p_+(\varphi)-\frac{p_-(\varphi)}{2}=\frac{p_{0,-}}{2h(\varphi)}\left\{\frac{m^2h^2(\varphi)+[\bm{p}_{0,\perp}-m\bm{\mathcal{F}}_{\perp}(\varphi)]^2}{p^2_{0,-}}-1\right\},\\
\label{epsilon}
\varepsilon(\varphi)&=p_+(\varphi)+\frac{p_-(\varphi)}{2}=\frac{p_{0,-}}{2h(\varphi)}\left\{\frac{m^2h^2(\varphi)+[\bm{p}_{0,\perp}-m\bm{\mathcal{F}}_{\perp}(\varphi)]^2}{p^2_{0,-}}+1\right\}.
\end{align}

Before replacing Eq. (\ref{iSol_A}) [or equivalently Eqs. (\ref{p_-})-(\ref{p_+})] in Eq. (\ref{dE_dodO_phase}), it is convenient to write the latter equation in the form
\begin{equation}
\label{dE_dodO_phase_f}
\frac{d\mathcal{E}}{d\bm{k}}=-\frac{e^2}{4\pi^2}\int d\varphi d\varphi'\frac{(p(\varphi)p(\varphi'))}{(k_0p(\varphi))(k_0p(\varphi'))}e^{i\int_{\varphi'}^{\varphi}d\tilde{\varphi}\frac{(kp(\tilde{\varphi}))}{(k_0p(\tilde{\varphi}))}},
\end{equation}
where we have introduced the four-wave-vector of the emitted radiation $k^{\mu}=(\omega,\bm{k})=\omega(1,\bm{n})$ and we have used the identity (see also Refs. \cite{Jackson_b_1975,Baier_b_1998} on this)
\begin{equation}
\int_{-\infty}^{\infty} d\varphi\frac{(kp(\varphi))}{(k_0p(\varphi))} e^{i\int_{-\infty}^{\varphi}d\tilde{\varphi}\frac{(kp(\tilde{\varphi}))}{(k_0p(\tilde{\varphi}))}}=0.
\end{equation}
Equation (\ref{dE_dodO_phase_f}) is especially useful if one expresses the four-dimensional scalar products in light-cone coordinates and exploits the fact that the electron four-momentum is on-shell, i.e., $p^2(\varphi)=m^2$. After a few straightforward manipulations, one can easily write Eq. (\ref{dE_dodO_phase_f}) in the form
\begin{equation}
\label{dE_dodO_phase_ff}
\frac{d\mathcal{E}}{d\bm{k}}=-\frac{e^2}{8\pi^2m^2\eta_0^2}\int d\varphi d\varphi'\,e^{i\frac{k_-}{2p_{0,-}\eta_0}\int_{\varphi'}^{\varphi}d\tilde{\varphi}h^2(\tilde{\varphi})[1+\bm{\pi}_{\perp}^2(\tilde{\varphi})]}\left\{h^2(\varphi)+h^2(\varphi')+[\bm{\mathcal{F}}_{\perp}(\varphi)-\bm{\mathcal{F}}_{\perp}(\varphi')]^2\right\},
\end{equation}
where
\begin{equation}
\label{pi}
\bm{\pi}_{\perp}(\varphi)=\frac{1}{m}\left[\bm{p}_{\perp}(\varphi)-\frac{p_-(\varphi)}{k_-}\bm{k}_{\perp}\right]=\frac{1}{mh(\varphi)}\left[\bm{p}_{0,\perp}-m\bm{\mathcal{F}}_{\perp}(\varphi)-\frac{p_{0,-}}{k_-}\bm{k}_{\perp}\right]
\end{equation}

This expression shows that the effects of radiation reaction are all encoded in the function $h(\varphi)$ [see Eq. (\ref{h})] and if radiation reaction is ignored, i.e., for $h(\varphi)=1$, one obtains the classical spectrum of Thomson scattering. This, in turn, can be obtained as the classical limit of the spectrum of nonlinear Compton scattering as reported, e.g., in Ref. \cite{Di_Piazza_2018}, which is accomplished by neglecting the recoil of the emitted radiation (emitted photon in the quantum language) on the electron. More precisely, we recall here that Eq. (\ref{dE_dodO_phase_ff}) divided by $\omega$ corresponds to the classical limit of the average number of photons emitted by the electron per units of emitted photon momentum \cite{Glauber_1951,Di_Piazza_2010,Tamburini_2019}.

As one can easily recognize, from Eq. (\ref{dE_dodO_phase_ff}) one can obtain the angularly-integrated energy emission spectrum $d\mathcal{E}/dk_-$ by using the fact that $d\bm{k}=(\omega/k_-)dk_-d\bm{k}_{\perp}$ and then
\begin{equation}
\begin{split}
\frac{d\mathcal{E}}{dk_-}&=-\frac{e^2}{8\pi^2m^2\eta_0^2}\int d\bm{k}_{\perp}\frac{\omega}{k_-}\int d\varphi d\varphi'\,e^{i\frac{k_-}{2p_{0,-}\eta_0}\int_{\varphi'}^{\varphi}d\tilde{\varphi}h^2(\tilde{\varphi})[1+\bm{\pi}_{\perp}^2(\tilde{\varphi})]}\\
&\quad\times \left\{h^2(\varphi)+h^2(\varphi')+[\bm{\mathcal{F}}_{\perp}(\varphi)-\bm{\mathcal{F}}_{\perp}(\varphi')]^2\right\}.
\end{split}
\end{equation}
By noticing that $\omega=k_++k_-/2=\bm{k}_{\perp}^2/2k_-+k_-/2$, the integral in $d\bm{k}_{\perp}$ is easily taken as it is Gaussian. By passing for convenience to the average and the relative phases $\varphi_+=(\varphi+\varphi')/2$ and $\varphi_-=\varphi-\varphi'$, the resulting energy spectrum is given by
\begin{equation}
\label{dE_do_phase_ff}
\begin{split}
\frac{d\mathcal{E}}{dk_-}&=-\frac{ie^2}{8\pi\eta_0}\frac{k_-}{p_{0,-}}\int \frac{d\varphi_+ d\varphi_-}{\varphi_-+i0}\,e^{i\frac{k_-}{2p_{0,-}\eta_0}\left\{\int_{-\varphi_-/2}^{\varphi_-/2}d\tilde{\varphi}[h^2(\varphi_++\tilde{\varphi})+\bm{\mathcal{F}}^2_{\perp}(\varphi_++\tilde{\varphi})]-\frac{1}{\varphi_-}\left[\int_{-\varphi_-/2}^{\varphi_-/2}d\tilde{\varphi}\,\bm{\mathcal{F}}_{\perp}(\varphi_++\tilde{\varphi})\right]^2\right\}}\\
&\quad\times \left\{h^2\left(\varphi_++\frac{\varphi_-}{2}\right)+h^2\left(\varphi_+-\frac{\varphi_-}{2}\right)+\left[\bm{\mathcal{F}}_{\perp}\left(\varphi_++\frac{\varphi_-}{2}\right)-\bm{\mathcal{F}}_{\perp}\left(\varphi_+-\frac{\varphi_-}{2}\right)\right]^2\right\}\\
&\quad\times\Bigglb( 1+\frac{m^2}{p_{0,-}^2}\left\{\frac{1}{\varphi_-}\int_{-\varphi_-/2}^{\varphi_-/2}d\tilde{\varphi}\left[\frac{\bm{p}_{0,\perp}}{m}-\bm{\mathcal{F}}_{\perp}(\varphi_++\tilde{\varphi})\right]\right\}^2+\frac{2im^2\eta_0}{k_-p_{0,-}}\frac{1}{\varphi_-+i0}\Bigglb),
\end{split}
\end{equation}
where the shift of the pole at $\varphi_-=0$ toward the negative imaginary half-plane can be understood by imposing that the Gaussian integral converges \cite{Baier_b_1998,Dinu_2016,Di_Piazza_2018}. 

We observe that the structure of the exponential function in the first line of this equation allows for introducing the concept of electron dressed mass inside a plane wave \cite{Brown_1964,Kibble_1975} also when radiation-reaction effects are important. Indeed, the phase-dependent electron square dressed mass $\tilde{m}^2(\varphi_+,\varphi_-)$ can be defined here as [see Eq. (\ref{dE_do_phase_ff})]
\begin{equation}
\begin{split}
\tilde{m}^2(\varphi_+,\varphi_-)
&=m^2\left\{\frac{1}{\varphi_-}\int_{-\varphi_-/2}^{\varphi_-/2}d\tilde{\varphi}\,h^2(\varphi_++\tilde{\varphi})\right.\\
&\left.\quad+\frac{1}{\varphi_-}\int_{-\varphi_-/2}^{\varphi_-/2}d\tilde{\varphi}\,\bm{\mathcal{F}}^2_{\perp}(\varphi_++\tilde{\varphi})-\left[\frac{1}{\varphi_-}\int_{-\varphi_-/2}^{\varphi_-/2}d\tilde{\varphi}\,\bm{\mathcal{F}}_{\perp}(\varphi_++\tilde{\varphi})\right]^2\right\}.
\end{split}
\end{equation}
This expression generalizes the phase-dependent square electron dressed mass as reported, e.g., in Refs. \cite{Brown_1964,Kibble_1975,Hebenstreit_2011,Di_Piazza_2018_d}, as it includes radiation-reaction effects. 

Equation (\ref{dE_do_phase_ff}) can be explicitly regularized. First, we integrate by parts the term proportional to $[h^2(\varphi_++\varphi_-/2)+h^2(\varphi_+-\varphi_-/2)]/\varphi_-^2$ and we obtain
\begin{equation}
\label{dE_do_phase_ff_r}
\begin{split}
\frac{d\mathcal{E}}{dk_-}&=-\frac{ie^2}{4\pi\eta_0}\frac{k_-}{p_{0,-}}\int \frac{d\varphi_+ d\varphi_-}{\varphi_-+i0}\,e^{i\frac{k_-}{2p_{0,-}}\frac{\tilde{m}^2(\varphi_+,\varphi_-)}{m^2\eta_0}\varphi_-}\\
&\quad\times\bigglb( \bar{h}_2(\varphi_+,\varphi_-)\left\{1+\frac{m^2}{p_{0,-}^2}[\bm{p}_{0,\perp}-\langle\bm{\mathcal{F}}_{\perp}\rangle(\varphi_+,\varphi_-)]^2-\frac{m^2}{p_{0,-}^2}\bar{h}_2(\varphi_+,\varphi_-)\right\}\\
&\quad +\frac{im^2\eta_0}{k_-p_{0,-}}\left[h\left(\varphi_++\frac{\varphi_-}{2}\right)h'\left(\varphi_++\frac{\varphi_-}{2}\right)-h\left(\varphi_+-\frac{\varphi_-}{2}\right)h'\left(\varphi_+-\frac{\varphi_-}{2}\right)\right]\\
&\quad-\frac{m^2}{p^2_{0,-}}\bar{h}_2(\varphi_+,\varphi_-)\left[\bar{\bm{\mathcal{F}}}_{2,\perp}(\varphi_+,\varphi_-)+\langle\bm{\mathcal{F}}_{\perp}\rangle^2(\varphi_+,\varphi_-)-2\bm{\bar{\mathcal{F}}}_{\perp}(\varphi_+,\varphi_-)\cdot\langle\bm{\mathcal{F}}_{\perp}\rangle(\varphi_+,\varphi_-)\right]\\
&\quad+\frac{1}{2}\left[\bm{\mathcal{F}}_{\perp}\left(\varphi_++\frac{\varphi_-}{2}\right)-\bm{\mathcal{F}}_{\perp}\left(\varphi_+-\frac{\varphi_-}{2}\right)\right]^2\\
&\quad\times\left\{1+\frac{m^2}{p_{0,-}^2}[\bm{p}_{0,\perp}-\langle\bm{\mathcal{F}}_{\perp}\rangle(\varphi_+,\varphi_-)]^2+\frac{2im^2\eta_0}{k_-p_{0,-}}\frac{1}{\varphi_-}\right\}\bigglb),
\end{split}
\end{equation}
where we have introduced the notation
\begin{align}
\bar{h}_2(\varphi_+,\varphi_-)&=\frac{1}{2}\left[h^2\left(\varphi_++\frac{\varphi_-}{2}\right)+h^2\left(\varphi_+-\frac{\varphi_-}{2}\right)\right],\\
\langle h^2\rangle(\varphi_+,\varphi_-)&=\frac{1}{\varphi_-}\int_{-\varphi_-/2}^{\varphi_-/2}d\tilde{\varphi}\,h^2(\varphi_++\tilde{\varphi}),\\
\bar{\bm{\mathcal{F}}}_{\perp}(\varphi_+,\varphi_-)&=\frac{1}{2}\left[\bm{\mathcal{F}}_{\perp}\left(\varphi_++\frac{\varphi_-}{2}\right)+\bm{\mathcal{F}}_{\perp}\left(\varphi_+-\frac{\varphi_-}{2}\right)\right],\\
\bar{\bm{\mathcal{F}}}_{2,\perp}(\varphi_+,\varphi_-)&=\frac{1}{2}\left[\bm{\mathcal{F}}^2_{\perp}\left(\varphi_++\frac{\varphi_-}{2}\right)+\bm{\mathcal{F}}^2_{\perp}\left(\varphi_+-\frac{\varphi_-}{2}\right)\right],\\
\langle\bm{\mathcal{F}}_{\perp}\rangle(\varphi_+,\varphi_-)&=\frac{1}{\varphi_-}\int_{-\varphi_-/2}^{\varphi_-/2}d\tilde{\varphi}\,\bm{\mathcal{F}}_{\perp}(\varphi_++\tilde{\varphi}),\\
\langle\bm{\mathcal{F}}^2_{\perp}\rangle(\varphi_+,\varphi_-)&=\frac{1}{\varphi_-}\int_{-\varphi_-/2}^{\varphi_-/2}d\tilde{\varphi}\,\bm{\mathcal{F}}^2_{\perp}(\varphi_++\tilde{\varphi}).
\end{align}
Note that with these definitions, the square of the electron dressed mass can be simply written as
\begin{equation}
\tilde{m}^2(\varphi_+,\varphi_-)=m^2\left[\langle h^2\rangle(\varphi_+,\varphi_-)+\langle\bm{\mathcal{F}}^2_{\perp}\rangle(\varphi_+,\varphi_-)-\langle\bm{\mathcal{F}}_{\perp}\rangle^2(\varphi_+,\varphi_-)\right].
\end{equation}
At this point only the terms in the second line of Eq. (\ref{dE_do_phase_ff_r}) need an explicit regularization. In the absence of radiation reaction, this is achieved by imposing that the emission spectrum has to vanish in the absence of the external field \cite{Baier_b_1998,Dinu_2016,Di_Piazza_2018}. Here, due to the effect of radiation reaction, we need to introduce a slightly more complicated regularization procedure. To this end, we introduce the function
\begin{equation}
H_2(\varphi_+,\varphi_-)=\varphi_-\langle h^2\rangle(\varphi_+,\varphi_-)=\int_{-\varphi_-/2}^{\varphi_-/2}d\tilde{\varphi}\,h^2(\varphi_++\tilde{\varphi})
\end{equation}
and notice that
\begin{equation}
\label{dH_dphim}
\frac{\partial H_2(\varphi_+,\varphi_-)}{\partial\varphi_-}=\bar{h}_2(\varphi_+,\varphi_-)>0
\end{equation}
for any $\varphi_+$. Now, for any positive real number $a$, it is
\begin{equation}
\int_{-\infty}^{\infty}\frac{dH_2}{H_2+i0}e^{iaH_2}=0.
\end{equation}
We have indicated the integration variable as $H_2$ here because, by exploiting the result in Eq. (\ref{dH_dphim}), we change variable to $\varphi_-$ and we obtain
\begin{equation}
\begin{split}
&\int_{-\infty}^{\infty}\frac{d\varphi_-}{H_2(\varphi_+,\varphi_-)+i0}\frac{\partial H_2(\varphi_+,\varphi_-)}{\partial\varphi_-}e^{iaH_2(\varphi_+,\varphi_-)}\\
&\quad=\int_{-\infty}^{\infty}\frac{d\varphi_-}{H_2(\varphi_+,\varphi_-)+i0}\bar{h}_2(\varphi_+,\varphi_-)e^{iaH_2(\varphi_+,\varphi_-)}=0.
\end{split}
\end{equation}
This result shows that we can formally regularize the remaining terms of Eq. (\ref{dE_do_phase_ff_r}) by subtracting the vanishing quantity
\begin{equation}
\begin{split}
&h^2(\varphi_+)\left\{1+\frac{m^2}{p_{0,-}^2}[\bm{p}_{0,\perp}-\bm{\mathcal{F}}_{\perp}(\varphi_+)]^2-\frac{m^2}{p_{0,-}^2}h^2(\varphi_+)\right\}\\
&\quad\times\int_{-\infty}^{\infty}\frac{d\varphi_-}{H_2(\varphi_+,\varphi_-)+i0}\bar{h}_2(\varphi_+,\varphi_-)e^{i\frac{k_-}{2p_{0,-}\eta_0}H_2(\varphi_+,\varphi_-)}
\end{split}
\end{equation}
inside the integral in $\varphi_+$. As it will be clear below, the additional front factor $h^2(\varphi_+)$ is included because for $|\varphi_-|\ll 1$ it is $H_2(\varphi_+,\varphi_-)\approx h^2(\varphi_+)\varphi_-$. The resulting regularized expression of the energy spectrum reads
\begin{equation}
\label{dE_do_phase_fff}
\begin{split}
&\frac{d\mathcal{E}}{dk_-}=-\frac{ie^2}{4\pi\eta_0}\frac{k_-}{p_{0,-}}\int d\varphi_+ d\varphi_-\,e^{i\frac{k_-}{2p_{0,-}}\frac{\tilde{m}^2(\varphi_+,\varphi_-)}{m^2\eta_0}\varphi_-}\\
&\quad\times\Bigglb( \bar{h}_2(\varphi_+,\varphi_-)\Bigg\{\frac{1+\frac{m^2}{p_{0,-}^2}[\bm{p}_{0,\perp}-\langle\bm{\mathcal{F}}_{\perp}\rangle(\varphi_+,\varphi_-)]^2-\frac{m^2}{p_{0,-}^2}\bar{h}_2(\varphi_+,\varphi_-)}{\varphi_-}\\
&\quad\left.-\frac{1+\frac{m^2}{p_{0,-}^2}[\bm{p}_{0,\perp}-\bm{\mathcal{F}}_{\perp}(\varphi_+)]^2-\frac{m^2}{p_{0,-}^2}h^2(\varphi_+)}{H_2(\varphi_+,\varphi_-)/h^2(\varphi_+)}e^{-i\frac{k_-}{2p_{0,-}\eta_0}\varphi_-\left[\langle\bm{\mathcal{F}}^2_{\perp}\rangle(\varphi_+,\varphi_-)-\langle\bm{\mathcal{F}}_{\perp}\rangle^2(\varphi_+,\varphi_-)\right]}\right\}\\
&\quad +\frac{im^2\eta_0}{k_-p_{0,-}}\frac{1}{\varphi_-}\left[h\left(\varphi_++\frac{\varphi_-}{2}\right)h'\left(\varphi_++\frac{\varphi_-}{2}\right)-h\left(\varphi_+-\frac{\varphi_-}{2}\right)h'\left(\varphi_+-\frac{\varphi_-}{2}\right)\right]\\
&\quad-\frac{m^2}{p^2_{0,-}}\frac{\bar{h}_2(\varphi_+,\varphi_-)}{\varphi_-}\left[\bar{\bm{\mathcal{F}}}_{2,\perp}(\varphi_+,\varphi_-)+\langle\bm{\mathcal{F}}_{\perp}\rangle^2(\varphi_+,\varphi_-)-2\bm{\bar{\mathcal{F}}}_{\perp}(\varphi_+,\varphi_-)\cdot\langle\bm{\mathcal{F}}_{\perp}\rangle(\varphi_+,\varphi_-)\right]\\
&\quad+\frac{1}{2\varphi_-}\left[\bm{\mathcal{F}}_{\perp}\left(\varphi_++\frac{\varphi_-}{2}\right)-\bm{\mathcal{F}}_{\perp}\left(\varphi_+-\frac{\varphi_-}{2}\right)\right]^2\\
&\quad\times\left\{1+\frac{m^2}{p_{0,-}^2}[\bm{p}_{0,\perp}-\langle\bm{\mathcal{F}}_{\perp}\rangle(\varphi_+,\varphi_-)]^2+\frac{2im^2\eta_0}{k_-p_{0,-}}\frac{1}{\varphi_-}\right\}\Bigglb)
,\end{split}
\end{equation}
where we have removed the now unnecessary shift $+i0$ of the pole. Notice that the above regularization prescription reduces to the known one in the absence of radiation reaction, which guarantees that the energy spectrum $d\mathcal{E}/dk_-$ vanishes if the external plane wave vanishes.

\section{The emission spectrum within the LCFA}
\label{LCFA}
In order to implement the LCFA, we use the same strategy as in Ref. \cite{Di_Piazza_2018} by expanding Eqs. (\ref{dE_dodO_phase_ff}) and (\ref{dE_do_phase_ff}) for small values of $|\varphi_-|$ [recall that within the LCFA the problematic term proportional to $1/(\varphi_-+i0)$ can be integrated analytically, see, e.g., Refs. \cite{Baier_b_1998,Di_Piazza_2018}, whereas it is easier to perform the integration by parts of the terms proportional to $1/(\varphi_-+i0)^2$ after the expansion for $|\varphi_-|\ll 1$]. 

It is interesting to notice that the regime where the LCFA applies well overlaps with the regime where classical radiation-reaction effects are large. In fact, the LCFA is typically applicable at large values of the classical nonlinearity parameter $\xi_0=|e|A_0/m=|e|E_0/m\omega_0$ \cite{Ritus_1985,Baier_b_1998,Di_Piazza_2012}, where $A_0=E_0/\omega_0$ and $E_0$ are the amplitude of the vector potential $\bm{A}_{\perp}(\varphi)$ and of the electric field $\bm{E}_{\perp}(\varphi)=-\omega_0\bm{A}'_{\perp}(\varphi)$ of the plane wave, such that $\xi_0$ is the amplitude of $\bm{\xi}_{\perp}(\varphi)=(e/m)\bm{A}'_{\perp}(\varphi)$ (see Refs. \cite{Baier_1989,Khokonov_2002,Di_Piazza_2007,Wistisen_2015,Harvey_2015,Dinu_2016,
Di_Piazza_2018,Blackburn_2018,Alexandrov_2019,Di_Piazza_2019, Ilderton_2019_b,Podszus_2019,Ilderton_2019,Lv_2021} for investigations about the limitations of the LCFA). Moreover, in the realm of classical electrodynamics one has to assume that the quantum nonlinearity parameter $\chi_0=\eta_0\xi_0$ is much smaller than unity \cite{Ritus_1985,Baier_b_1998,Di_Piazza_2012}. Under these conditions the LCFA is expected to be very accurate except possibly for extremely small emitted radiation frequencies, which we do not consider here \cite{Di_Piazza_2018,Di_Piazza_2019, Ilderton_2019_b}. This indeed well overlaps with the regime where classical radiation-reaction effects are typically large because, apart from long laser pulses, radiation-reaction effects become large for $\xi_0\gg 1$ [see Eq. (\ref{h})] but still with $\chi_0\ll 1$, to be able to neglect quantum corrections.

Under the above assumptions, as it is known, one has to expand the phases in Eqs. (\ref{dE_dodO_phase_ff}) and (\ref{dE_do_phase_ff}) up to the third order in $\varphi_-$, whereas the leading-order expansion is sufficient for the preexponential functions. The resulting angularly-resolved and angularly-integrated energy spectra within the LCFA can be written as [see the appendix for details on the more involved derivation of Eq. (\ref{dE_dk-_LCFA})]
\begin{align}
\label{dE_d3k_LCFA}
\begin{split}
\frac{d\mathcal{E}_{\text{LCFA}}}{d\bm{k}}&=\frac{e^2}{\sqrt{3}\pi^2}\frac{1}{m^2\eta_0}\int d\varphi_+\frac{h^2(\varphi_+)}{\chi(\varphi_+)}\sqrt{1+\bm{\pi}^2_{\perp}(\varphi_+)}\\
&\quad\times\left[1+2\bm{\pi}^2_{\perp}(\varphi_+)\right]\text{K}_{1/3}\left(\frac{2}{3}\frac{k_-}{p_{0,-}}\frac{h^2(\varphi_+)}{\chi(\varphi_+)}\left[1+\bm{\pi}^2_{\perp}(\varphi_+)\right]^{3/2}\right),
\end{split}\\
\label{dE_dk-_LCFA}
\begin{split}
\frac{d\mathcal{E}_{\text{LCFA}}}{dk_-}&=\frac{2e^2}{\sqrt{3}\pi}\frac{k_-}{p_{0,-}}\int d\varphi_+\frac{\varepsilon(\varphi_+)}{p_-(\varphi_+)}\frac{h^2(\varphi_+)}{\eta_0}\\
&\quad\times\left[\text{K}_{2/3}\left(\frac{2}{3}\frac{k_-}{p_{0,-}}\frac{h^2(\varphi_+)}{\chi(\varphi_+)}\right)-\frac{1}{2}\text{IK}_{1/3}\left(\frac{2}{3}\frac{k_-}{p_{0,-}}\frac{h^2(\varphi_+)}{\chi(\varphi_+)}\right)\right].
\end{split}
\end{align}
Here, we have introduced the local quantum nonlinearity parameter $\chi(\varphi)=\eta_0|\bm{\xi}_{\perp}(\varphi)|$ (it is easily checked that the above formulas do not contain $\hbar$ explicitly), the modified Bessel function $\text{K}_{\nu}(z)$ of order $\nu$ \cite{NIST_b_2010} and the function
\begin{equation}
\text{IK}_{\nu}(z)=\int_z^{\infty}dz'\text{K}_{\nu}(z').
\end{equation}
As expected from the very meaning of the LCFA, the above Eqs. (\ref{dE_d3k_LCFA})-(\ref{dE_dk-_LCFA}) can be obtained from the corresponding expressions in the absence of radiation reaction by replacing the components of the electron four-momentum obtained from solving the Lorentz equation in the plane wave with the corresponding expressions obtained from solving the LL equation [see Eqs. (\ref{p_-})-(\ref{epsilon})]. In particular, one can find that in the absence of radiation reaction Eq. (\ref{dE_dk-_LCFA}) has exactly the same form as the classical limit of the quantum energy emitted spectrum as computed in Ref. \cite{Ritus_1985}. However, we point out that the quantities $d\mathcal{E}_{\text{LCFA}}/d\bm{k}$ and $d\mathcal{E}_{\text{LCFA}}/dk_-$ are not local in $\varphi_+$ because both the function $h(\varphi)$ [see Eq. (\ref{h})] and the function $\bm{\mathcal{F}}_{\perp}(\varphi)$ [see the definitions below Eqs. (\ref{F_munu}) and (\ref{p_+})] are not local in the laser phase. This is also expected from the physical meaning of radiation reaction, with one of the main physical consequences being the accumulation effects of energy-momentum loss.

As an additional remark, we notice that by taking the integral of Eq. (\ref{dE_dk-_LCFA}) in $dk_-$ one obtains that the total energy radiated is given by
\begin{equation}
\mathcal{E}_{\text{LCFA}}=\frac{2}{3}e^2\eta_0\int d\varphi_+\frac{\varepsilon(\varphi_+)}{h(\varphi_+)}\bm{\xi}_{\perp}^2(\varphi_+).
\end{equation}
Although this result is obtained within the LCFA, it coincides with the total energy radiated by the electron in general, i.e., also \emph{beyond} the LCFA [see Eq. (19) in Ref. \cite{Di_Piazza_2018_b}]. This circumstance also occurs in the absence of radiation reaction, as it can be ascertained by comparing the general relativistic Larmor formula $d\mathcal{E}/dt=-(2/3)e^2a^2=(2/3)e^2m^2\chi^2$ (see, e.g., Ref. \cite{Jackson_b_1975}), where $a^{\mu}=\mathcal{F}^{\mu\nu}u_{\nu}/m$ is the four-acceleration of the electron, with the last equation on page 522 in Ref. \cite{Ritus_1985}.

Interestingly, the total minus component
\begin{equation}
\label{dK_-_dphi}
\frac{d\mathcal{K}_-}{d\varphi_+}=\int_0^{\infty} dk_-\int d\bm{k}_{\perp}\frac{k_-}{\omega}\frac{d\mathcal{E}}{dk_-d\bm{k}_{\perp}d\varphi_+}=\int_0^{\infty} dk_-\int d\bm{k}_{\perp}\frac{d\mathcal{E}}{d\bm{k}d\varphi_+}
\end{equation}
of the four-momentum radiated classically per unit of laser phase by an electron in a plane wave including radiation reaction has been recently computed within the LCFA in Ref. \cite{Heinzl_2021} in the different context of the so-called Ritus-Narozhny conjecture on strong-field QED \cite{Ritus_1970,Narozhny_1979,Narozhny_1980,Morozov_1981,Akhmedov_2011,Fedotov_2017}. According to Eq. (\ref{dK_-_dphi}), by defining $d\mathcal{E}_{\text{LCFA}}/d\bm{k}d\varphi_+$ as the integrand in Eq. (\ref{dE_d3k_LCFA}), and by performing the integral of this quantity over $d\bm{k}_{\perp}$ one can easily show that
\begin{equation}
\label{dK-_dk-_LCFA}
\begin{split}
\frac{d\mathcal{K}_{-,\text{LCFA}}}{d\varphi_+}&=\frac{2e^2}{\sqrt{3}\pi}\int_0^{\infty} dk_-\frac{k_-}{p_{0,-}}\frac{h^2(\varphi_+)}{\eta_0}\left[\text{K}_{2/3}\left(\frac{2}{3}\frac{k_-}{p_{0,-}}\frac{h^2(\varphi_+)}{\chi(\varphi_+)}\right)-\frac{1}{2}\text{IK}_{1/3}\left(\frac{2}{3}\frac{k_-}{p_{0,-}}\frac{h^2(\varphi_+)}{\chi(\varphi_+)}\right)\right]
\end{split}
\end{equation}
in agreement with the result in Ref. \cite{Heinzl_2021}.

\section{Conclusions}

In conclusion, we have derived analytically the angularly-resolved and the angularly-integrated energy emission spectra of nonlinear Thomson scattering by including radiation-reaction effects. This has been accomplished by starting from the analytical solution of the LL in an arbitrary plane wave and by using the classical formulas of radiation by accelerated charges. 

The general expressions of the spectra are presented in Eqs. (\ref{dE_dodO_phase_ff}) and (\ref{dE_do_phase_fff}), are valid for an arbitrary plane wave, and are obtained as double integrals over the plane-wave phase. A particular, new regularization technique has to be used in order to regularize the angularly-integrated spectrum. We point out that the resulting spectra include higher-order classical radiative corrections according to the Landau and Lifshitz reduction of order, meaning that neglected classical corrections are much smaller than quantum corrections, which have been of course ignored from the beginning.

Moreover, we have obtained a phase-dependent expression of the electron dressed mass, which includes radiation-reaction effects.

Finally, the expressions of the angularly-resolved and the angularly-integrated spectra within the locally constant field approximations have been derived as well. These expressions have the property that are expressed as single integrals over the laser phase of the corresponding expressions without radiation reaction with the electron four-momentum replaced with its expression including radiation reaction. Thus,  they turn out to be nonlocal exactly for the nature itself of radiation reaction giving rise to cumulative energy-momentum loss effects. 

\appendix
\setcounter{secnumdepth}{0}
\section{Appendix: Derivation of Eq. (\ref{dE_dk-_LCFA})}

The staring point is Eq. (\ref{dE_do_phase_ff}) and in order to implement the LCFA there, we expand each term of the preexponent up to the leading order for $|\varphi_-|\ll 1$, whereas we keep terms up to $\varphi_-^3$ in the phase (see, e.g., \cite{Di_Piazza_2018}):
\begin{equation}
\begin{split}
\frac{d\mathcal{E}_{\text{LCFA}}}{dk_-}&=-\frac{ie^2}{4\pi\eta_0}\frac{k_-}{p_{0,-}}\int d\varphi_+\,h^2(\varphi_+)\int \frac{d\varphi_-}{\varphi_-+i0}\,e^{i\frac{k_-}{2p_{0,-}\eta_0}h^2(\varphi_+)\varphi_-\left[1+\frac{1}{12}\bm{\xi}^2_{\perp}(\varphi_+)\varphi_-^2\right]}\\
&\quad\times \left[1+\frac{1}{2}\bm{\xi}^2_{\perp}(\varphi_+)\varphi_-^2\right]\left\{1+\frac{m^2}{p_{0,-}^2}\left[\frac{\bm{p}_{0,\perp}}{m}-\bm{\mathcal{F}}_{\perp}(\varphi_+)\right]^2+\frac{2im^2\eta_0}{k_-p_{0,-}}\frac{1}{\varphi_-+i0}\right\}.
\end{split}
\end{equation}
Now, we integrate by parts the only term containing $1/(\varphi_-+i0)^2$ in the preexponent and we obtain
\begin{equation}
\begin{split}
\frac{d\mathcal{E}_{\text{LCFA}}}{dk_-}&=-\frac{ie^2}{4\pi\eta_0}\frac{k_-}{p_{0,-}}\int d\varphi_+\,h^2(\varphi_+)\int \frac{d\varphi_-}{\varphi_-+i0}\,e^{i\frac{k_-}{2p_{0,-}\eta_0}h^2(\varphi_+)\varphi_-\left[1+\frac{1}{12}\bm{\xi}^2_{\perp}(\varphi_+)\varphi_-^2\right]}\\
&\quad\times \bigglb ( \left\{1+\frac{m^2}{p_{0,-}^2}\left[\frac{\bm{p}_{0,\perp}}{m}-\bm{\mathcal{F}}_{\perp}(\varphi_+)\right]^2\right\}\left[1+\frac{1}{2}\bm{\xi}^2_{\perp}(\varphi_+)\varphi_-^2\right]\\
&\qquad-\frac{m^2}{p_{0,-}^2}h^2(\varphi_+)\left[1+\frac{1}{4}\bm{\xi}^2_{\perp}(\varphi_+)\varphi_-^2\right]+\frac{im^2\eta_0}{k_-p_{0,-}}\bm{\xi}^2_{\perp}(\varphi_+)\varphi_-\bigglb).
\end{split}
\end{equation}
This equation is already regular and can been expressed in terms of modified Bessel functions but, for the sake of convenience, we integrate by parts the last term and have
\begin{equation}
\label{dE_dk-_LCFA_d}
\begin{split}
\frac{d\mathcal{E}_{\text{LCFA}}}{dk_-}&=-\frac{ie^2}{4\pi\eta_0}\frac{k_-}{p_{0,-}}\int d\varphi_+\,h^2(\varphi_+)\int \frac{d\varphi_-}{\varphi_-+i0}\,e^{i\frac{k_-}{2p_{0,-}\eta_0}h^2(\varphi_+)\varphi_-\left[1+\frac{1}{12}\bm{\xi}^2_{\perp}(\varphi_+)\varphi_-^2\right]}\\
&\quad\times \bigglb(\left\{1+\frac{m^2}{p_{0,-}^2}\left[\frac{\bm{p}_{0,\perp}}{m}-\bm{\mathcal{F}}_{\perp}(\varphi_+)\right]^2\right\}\left[1+\frac{1}{2}\bm{\xi}^2_{\perp}(\varphi_+)\varphi_-^2\right]\\
&\qquad-\frac{m^2}{p_{0,-}^2}h^2(\varphi_+)\left[1-\frac{1}{2}\bm{\xi}^2_{\perp}(\varphi_+)\varphi_-^2\right]\left[1+\frac{1}{4}\bm{\xi}^2_{\perp}(\varphi_+)\varphi_-^2\right]\bigglb)\\
&=-\frac{ie^2}{4\pi\eta_0}\frac{k_-}{p_{0,-}}\int d\varphi_+\,h^2(\varphi_+)\int \frac{dy}{y+i0}\,e^{i\frac{k_-}{p_{0,-}}\frac{h^2(\varphi_+)}{\chi(\varphi_+)}y\left(1+\frac{y^2}{3}\right)}\\
&\quad\times(1+2y^2)\left\{1+\frac{m^2}{p_{0,-}^2}\left[\frac{\bm{p}_{0,\perp}}{m}-\bm{\mathcal{F}}_{\perp}(\varphi_+)\right]^2-\frac{m^2}{p_{0,-}^2}h^2(\varphi_+)\frac{1-y^2-2y^4}{1+2y^2}\right\},
\end{split}
\end{equation}
where we have introduced local quantum nonlinearity parameter $\chi(\varphi)=\eta_0|\bm{\xi}_{\perp}(\varphi)|$ (see also the main text). At this point, we observe that the main contribution to the integral in $y=\varphi_-|\bm{\xi}_{\perp}(\varphi)|/2$ comes from the region $|y|\lesssim 1$. Moreover, we recall that within the LCFA we are assuming that $\xi_0\gg 1$ (see the discussion at the beginning of Sec. \ref{LCFA}), which means that the largest contribution to the integral in $\varphi_+$ comes from the regions where $\bm{\mathcal{F}}_{\perp}(\varphi_+)$ is at the largest. From the definitions below Eqs. (\ref{F_munu}) and (\ref{p_+}), we obtain that
\begin{equation}
\begin{split}
\bm{\mathcal{F}}_{\perp}(\varphi)&=\int_{-\infty}^{\varphi}d\tilde{\varphi}\,h(\tilde{\varphi})\bm{\xi}_{\perp}(\tilde{\varphi})=\frac{e}{m}\int_{-\infty}^{\varphi}d\tilde{\varphi}\,h(\tilde{\varphi})\bm{A}'_{\perp}(\varphi)\\
&=\frac{e}{m}\left[h(\varphi)\bm{A}_{\perp}(\varphi)-\frac{2}{3}e^2\eta_0\int_{-\infty}^{\varphi}d\tilde{\varphi}\,\bm{\xi}^2_{\perp}(\tilde{\varphi})\bm{A}_{\perp}(\tilde{\varphi})\right],
\end{split}
\end{equation}
which shows that $|\bm{\mathcal{F}}_{\perp}(\varphi)|\lesssim h(\varphi)\xi_0$. In conclusion, we can consistently neglect the last term in Eq. (\ref{dE_dk-_LCFA_d}) as compared to the second-last one within the LCFA (note that we do not make any assumptions about the values of $|\bm{p}_{0,\perp}|/m$ and $p_{0,-}/m$ as compared with $\xi_0$) and we finally obtain the expression in the main text:
\begin{equation}
\label{dE_dk-_LCFA_df}
\begin{split}
\frac{d\mathcal{E}_{\text{LCFA}}}{dk_-}&=\frac{2e^2}{\sqrt{3}\pi}\frac{k_-}{p_{0,-}}\int d\varphi_+\,\frac{\varepsilon(\varphi_+)}{p_-(\varphi_+)}\frac{h^2(\varphi_+)}{\eta_0}\\
&\quad\times\left[\text{K}_{2/3}\left(\frac{2}{3}\frac{k_-}{p_{0,-}}\frac{h^2(\varphi_+)}{\chi(\varphi_+)}\right)-\frac{1}{2}\text{IK}_{1/3}\left(\frac{2}{3}\frac{k_-}{p_{0,-}}\frac{h^2(\varphi_+)}{\chi(\varphi_+)}\right)\right],
\end{split}
\end{equation}
where we have used the integral definitions of the modified Bessel functions $K_{\nu}(z)$ \cite{NIST_b_2010} and the expression (\ref{epsilon}) of the energy of the electron inside the plane wave, neglecting the term proportional to $m^2$ there. The last approximations used prevent the possibility of interpreting the integrand of Eq. (\ref{dE_dk-_LCFA_df}) as the energy emitted per unit of laser phase and unit of $k_-$.


%\bibliography{g://Travagghiu/Bibliography/arXiv,g://Travagghiu/Bibliography/Books,g://Travagghiu/Bibliography/Reviews,g://Travagghiu/Bibliography/Papers_Nuclei,g://Travagghiu/Bibliography/Papers_Radiation,g://Travagghiu/Bibliography/Papers_RR,g://Travagghiu/Bibliography/Papers_PP_and_Cascades,g://Travagghiu/Bibliography/Papers_VPE,g://Travagghiu/Bibliography/Papers_Crystal,g://Travagghiu/Bibliography/Papers_Various,g://Travagghiu/Bibliography/Papers_Laser_Plasma_QED,g://Travagghiu/Bibliography/Papers_XFEL,g://Travagghiu/Bibliography/Homepages}

\begin{thebibliography}{65}%
\makeatletter
\providecommand \@ifxundefined [1]{%
 \@ifx{#1\undefined}
}%
\providecommand \@ifnum [1]{%
 \ifnum #1\expandafter \@firstoftwo
 \else \expandafter \@secondoftwo
 \fi
}%
\providecommand \@ifx [1]{%
 \ifx #1\expandafter \@firstoftwo
 \else \expandafter \@secondoftwo
 \fi
}%
\providecommand \natexlab [1]{#1}%
\providecommand \enquote  [1]{``#1''}%
\providecommand \bibnamefont  [1]{#1}%
\providecommand \bibfnamefont [1]{#1}%
\providecommand \citenamefont [1]{#1}%
\providecommand \href@noop [0]{\@secondoftwo}%
\providecommand \href [0]{\begingroup \@sanitize@url \@href}%
\providecommand \@href[1]{\@@startlink{#1}\@@href}%
\providecommand \@@href[1]{\endgroup#1\@@endlink}%
\providecommand \@sanitize@url [0]{\catcode `\\12\catcode `\$12\catcode
  `\&12\catcode `\#12\catcode `\^12\catcode `\_12\catcode `\%12\relax}%
\providecommand \@@startlink[1]{}%
\providecommand \@@endlink[0]{}%
\providecommand \url  [0]{\begingroup\@sanitize@url \@url }%
\providecommand \@url [1]{\endgroup\@href {#1}{\urlprefix }}%
\providecommand \urlprefix  [0]{URL }%
\providecommand \Eprint [0]{\href }%
\providecommand \doibase [0]{https://doi.org/}%
\providecommand \selectlanguage [0]{\@gobble}%
\providecommand \bibinfo  [0]{\@secondoftwo}%
\providecommand \bibfield  [0]{\@secondoftwo}%
\providecommand \translation [1]{[#1]}%
\providecommand \BibitemOpen [0]{}%
\providecommand \bibitemStop [0]{}%
\providecommand \bibitemNoStop [0]{.\EOS\space}%
\providecommand \EOS [0]{\spacefactor3000\relax}%
\providecommand \BibitemShut  [1]{\csname bibitem#1\endcsname}%
\let\auto@bib@innerbib\@empty
%</preamble>
\bibitem [{\citenamefont {Jackson}(1975)}]{Jackson_b_1975}%
  \BibitemOpen
  \bibfield  {author} {\bibinfo {author} {\bibfnamefont {J.~D.}\ \bibnamefont
  {Jackson}},\ }\href@noop {} {\emph {\bibinfo {title} {Classical
  Electrodynamics}}}\ (\bibinfo  {publisher} {John Wiley \& Sons, New York},\
  \bibinfo {year} {1975})\BibitemShut {NoStop}%
\bibitem [{\citenamefont {Landau}\ and\ \citenamefont
  {Lifshitz}(1975)}]{Landau_b_2_1975}%
  \BibitemOpen
  \bibfield  {author} {\bibinfo {author} {\bibfnamefont {L.~D.}\ \bibnamefont
  {Landau}}\ and\ \bibinfo {author} {\bibfnamefont {E.~M.}\ \bibnamefont
  {Lifshitz}},\ }\href@noop {} {\emph {\bibinfo {title} {The Classical Theory
  of Fields}}}\ (\bibinfo  {publisher} {Elsevier, Oxford},\ \bibinfo {year}
  {1975})\BibitemShut {NoStop}%
\bibitem [{\citenamefont {Barut}(1980)}]{Barut_b_1980}%
  \BibitemOpen
  \bibfield  {author} {\bibinfo {author} {\bibfnamefont {A.~O.}\ \bibnamefont
  {Barut}},\ }\href@noop {} {\emph {\bibinfo {title} {Electrodynamics and
  Classical Theory of Fields and Particles}}}\ (\bibinfo  {publisher} {Dover,
  New York},\ \bibinfo {year} {1980})\BibitemShut {NoStop}%
\bibitem [{\citenamefont {Rohrlich}(2007)}]{Rohrlich_b_2007}%
  \BibitemOpen
  \bibfield  {author} {\bibinfo {author} {\bibfnamefont {F.}~\bibnamefont
  {Rohrlich}},\ }\href@noop {} {\emph {\bibinfo {title} {Classical Charged
  Particles}}}\ (\bibinfo  {publisher} {World Scientific, Singapore},\ \bibinfo
  {year} {2007})\BibitemShut {NoStop}%
\bibitem [{\citenamefont {Abraham}(1905)}]{Abraham_b_1905}%
  \BibitemOpen
  \bibfield  {author} {\bibinfo {author} {\bibfnamefont {M.}~\bibnamefont
  {Abraham}},\ }\href@noop {} {\emph {\bibinfo {title} {Theorie der
  Elektrizit{\"a}t}}}\ (\bibinfo  {publisher} {Teubner, Leipzig},\ \bibinfo
  {year} {1905})\BibitemShut {NoStop}%
\bibitem [{\citenamefont {Lorentz}(1909)}]{Lorentz_b_1909}%
  \BibitemOpen
  \bibfield  {author} {\bibinfo {author} {\bibfnamefont {H.~A.}\ \bibnamefont
  {Lorentz}},\ }\href@noop {} {\emph {\bibinfo {title} {The Theory of
  Electrons}}}\ (\bibinfo  {publisher} {Teubner, Leipzig},\ \bibinfo {year}
  {1909})\BibitemShut {NoStop}%
\bibitem [{\citenamefont {Dirac}(1938)}]{Dirac_1938}%
  \BibitemOpen
  \bibfield  {author} {\bibinfo {author} {\bibfnamefont {P.~A.~M.}\
  \bibnamefont {Dirac}},\ }\href@noop {} {\bibfield  {journal} {\bibinfo
  {journal} {Proc. R. Soc. London, Ser. A}\ }\textbf {\bibinfo {volume}
  {167}},\ \bibinfo {pages} {148} (\bibinfo {year} {1938})}\BibitemShut
  {NoStop}%
\bibitem [{\citenamefont {Spohn}(2000)}]{Spohn_2000}%
  \BibitemOpen
  \bibfield  {author} {\bibinfo {author} {\bibfnamefont {H.}~\bibnamefont
  {Spohn}},\ }\href@noop {} {\bibfield  {journal} {\bibinfo  {journal}
  {Europhys. Lett.}\ }\textbf {\bibinfo {volume} {50}},\ \bibinfo {pages} {287}
  (\bibinfo {year} {2000})}\BibitemShut {NoStop}%
\bibitem [{\citenamefont {Koga}(2004)}]{Koga_2004}%
  \BibitemOpen
  \bibfield  {author} {\bibinfo {author} {\bibfnamefont {J.}~\bibnamefont
  {Koga}},\ }\href@noop {} {\bibfield  {journal} {\bibinfo  {journal} {Phys.
  Rev. E}\ }\textbf {\bibinfo {volume} {70}},\ \bibinfo {pages} {046502}
  (\bibinfo {year} {2004})}\BibitemShut {NoStop}%
\bibitem [{\citenamefont {Bulanov}\ \emph {et~al.}(2011)\citenamefont
  {Bulanov}, \citenamefont {Esirkepov}, \citenamefont {Kando}, \citenamefont
  {Koga},\ and\ \citenamefont {Bulanov}}]{Bulanov_2011}%
  \BibitemOpen
  \bibfield  {author} {\bibinfo {author} {\bibfnamefont {S.~V.}\ \bibnamefont
  {Bulanov}}, \bibinfo {author} {\bibfnamefont {T.~Z.}\ \bibnamefont
  {Esirkepov}}, \bibinfo {author} {\bibfnamefont {M.}~\bibnamefont {Kando}},
  \bibinfo {author} {\bibfnamefont {J.~K.}\ \bibnamefont {Koga}},\ and\
  \bibinfo {author} {\bibfnamefont {S.~S.}\ \bibnamefont {Bulanov}},\
  }\href@noop {} {\bibfield  {journal} {\bibinfo  {journal} {Phys. Rev. E}\
  }\textbf {\bibinfo {volume} {84}},\ \bibinfo {pages} {056605} (\bibinfo
  {year} {2011})}\BibitemShut {NoStop}%
\bibitem [{\citenamefont {Hadad}\ \emph {et~al.}(2010)\citenamefont {Hadad},
  \citenamefont {Labun}, \citenamefont {Rafelski}, \citenamefont {Elkina},
  \citenamefont {Klier},\ and\ \citenamefont {Ruhl}}]{Hadad_2010}%
  \BibitemOpen
  \bibfield  {author} {\bibinfo {author} {\bibfnamefont {Y.}~\bibnamefont
  {Hadad}}, \bibinfo {author} {\bibfnamefont {L.}~\bibnamefont {Labun}},
  \bibinfo {author} {\bibfnamefont {J.}~\bibnamefont {Rafelski}}, \bibinfo
  {author} {\bibfnamefont {N.}~\bibnamefont {Elkina}}, \bibinfo {author}
  {\bibfnamefont {C.}~\bibnamefont {Klier}},\ and\ \bibinfo {author}
  {\bibfnamefont {H.}~\bibnamefont {Ruhl}},\ }\href
  {https://doi.org/10.1103/PhysRevD.82.096012} {\bibfield  {journal} {\bibinfo
  {journal} {Phys. Rev. D}\ }\textbf {\bibinfo {volume} {82}},\ \bibinfo
  {pages} {096012} (\bibinfo {year} {2010})}\BibitemShut {NoStop}%
\bibitem [{\citenamefont {Vranic}\ \emph {et~al.}(2014)\citenamefont {Vranic},
  \citenamefont {Martins}, \citenamefont {Vieira}, \citenamefont {Fonseca},\
  and\ \citenamefont {Silva}}]{Vranic_2014}%
  \BibitemOpen
  \bibfield  {author} {\bibinfo {author} {\bibfnamefont {M.}~\bibnamefont
  {Vranic}}, \bibinfo {author} {\bibfnamefont {J.~L.}\ \bibnamefont {Martins}},
  \bibinfo {author} {\bibfnamefont {J.}~\bibnamefont {Vieira}}, \bibinfo
  {author} {\bibfnamefont {R.~A.}\ \bibnamefont {Fonseca}},\ and\ \bibinfo
  {author} {\bibfnamefont {L.~O.}\ \bibnamefont {Silva}},\ }\href@noop {}
  {\bibfield  {journal} {\bibinfo  {journal} {Phys. Rev. Lett.}\ }\textbf
  {\bibinfo {volume} {113}},\ \bibinfo {pages} {134801} (\bibinfo {year}
  {2014})}\BibitemShut {NoStop}%
\bibitem [{\citenamefont {Blackburn}\ \emph {et~al.}(2014)\citenamefont
  {Blackburn}, \citenamefont {Ridgers}, \citenamefont {Kirk},\ and\
  \citenamefont {Bell}}]{Blackburn_2014}%
  \BibitemOpen
  \bibfield  {author} {\bibinfo {author} {\bibfnamefont {T.~G.}\ \bibnamefont
  {Blackburn}}, \bibinfo {author} {\bibfnamefont {C.~P.}\ \bibnamefont
  {Ridgers}}, \bibinfo {author} {\bibfnamefont {J.~G.}\ \bibnamefont {Kirk}},\
  and\ \bibinfo {author} {\bibfnamefont {A.~R.}\ \bibnamefont {Bell}},\
  }\href@noop {} {\bibfield  {journal} {\bibinfo  {journal} {Phys. Rev. Lett.}\
  }\textbf {\bibinfo {volume} {112}},\ \bibinfo {pages} {015001} (\bibinfo
  {year} {2014})}\BibitemShut {NoStop}%
\bibitem [{\citenamefont {Tamburini}\ \emph {et~al.}(2014)\citenamefont
  {Tamburini}, \citenamefont {Keitel},\ and\ \citenamefont
  {Di~Piazza}}]{Tamburini_2014}%
  \BibitemOpen
  \bibfield  {author} {\bibinfo {author} {\bibfnamefont {M.}~\bibnamefont
  {Tamburini}}, \bibinfo {author} {\bibfnamefont {C.~H.}\ \bibnamefont
  {Keitel}},\ and\ \bibinfo {author} {\bibfnamefont {A.}~\bibnamefont
  {Di~Piazza}},\ }\href@noop {} {\bibfield  {journal} {\bibinfo  {journal}
  {Phys. Rev. E}\ }\textbf {\bibinfo {volume} {89}},\ \bibinfo {pages}
  {021201(R)} (\bibinfo {year} {2014})}\BibitemShut {NoStop}%
\bibitem [{\citenamefont {Li}\ \emph {et~al.}(2014)\citenamefont {Li},
  \citenamefont {Hatsagortsyan},\ and\ \citenamefont {Keitel}}]{Li_2014}%
  \BibitemOpen
  \bibfield  {author} {\bibinfo {author} {\bibfnamefont {J.-X.}\ \bibnamefont
  {Li}}, \bibinfo {author} {\bibfnamefont {K.~Z.}\ \bibnamefont
  {Hatsagortsyan}},\ and\ \bibinfo {author} {\bibfnamefont {C.~H.}\
  \bibnamefont {Keitel}},\ }\href@noop {} {\bibfield  {journal} {\bibinfo
  {journal} {Phys. Rev. Lett.}\ }\textbf {\bibinfo {volume} {113}},\ \bibinfo
  {pages} {044801} (\bibinfo {year} {2014})}\BibitemShut {NoStop}%
\bibitem [{\citenamefont {Heinzl}\ \emph {et~al.}(2015)\citenamefont {Heinzl},
  \citenamefont {Harvey}, \citenamefont {Ilderton}, \citenamefont {Marklund},
  \citenamefont {Bulanov}, \citenamefont {Rykovanov}, \citenamefont
  {Schroeder}, \citenamefont {Esarey},\ and\ \citenamefont
  {Leemans}}]{Heinzl_2015}%
  \BibitemOpen
  \bibfield  {author} {\bibinfo {author} {\bibfnamefont {T.}~\bibnamefont
  {Heinzl}}, \bibinfo {author} {\bibfnamefont {C.}~\bibnamefont {Harvey}},
  \bibinfo {author} {\bibfnamefont {A.}~\bibnamefont {Ilderton}}, \bibinfo
  {author} {\bibfnamefont {M.}~\bibnamefont {Marklund}}, \bibinfo {author}
  {\bibfnamefont {S.~S.}\ \bibnamefont {Bulanov}}, \bibinfo {author}
  {\bibfnamefont {S.}~\bibnamefont {Rykovanov}}, \bibinfo {author}
  {\bibfnamefont {C.~B.}\ \bibnamefont {Schroeder}}, \bibinfo {author}
  {\bibfnamefont {E.}~\bibnamefont {Esarey}},\ and\ \bibinfo {author}
  {\bibfnamefont {W.~P.}\ \bibnamefont {Leemans}},\ }\href@noop {} {\bibfield
  {journal} {\bibinfo  {journal} {Phys. Rev. E}\ }\textbf {\bibinfo {volume}
  {91}},\ \bibinfo {pages} {023207} (\bibinfo {year} {2015})}\BibitemShut
  {NoStop}%
\bibitem [{\citenamefont {Yoffe}\ \emph {et~al.}(2015)\citenamefont {Yoffe},
  \citenamefont {Kravets}, \citenamefont {Noble},\ and\ \citenamefont
  {Jaroszynski}}]{Yoffe_2015}%
  \BibitemOpen
  \bibfield  {author} {\bibinfo {author} {\bibfnamefont {S.~R.}\ \bibnamefont
  {Yoffe}}, \bibinfo {author} {\bibfnamefont {Y.}~\bibnamefont {Kravets}},
  \bibinfo {author} {\bibfnamefont {A.}~\bibnamefont {Noble}},\ and\ \bibinfo
  {author} {\bibfnamefont {D.~A.}\ \bibnamefont {Jaroszynski}},\ }\href@noop {}
  {\bibfield  {journal} {\bibinfo  {journal} {New J. Phys.}\ }\textbf {\bibinfo
  {volume} {17}},\ \bibinfo {pages} {053025} (\bibinfo {year}
  {2015})}\BibitemShut {NoStop}%
\bibitem [{\citenamefont {Capdessus}\ and\ \citenamefont
  {McKenna}(2015)}]{Capdessus_2015}%
  \BibitemOpen
  \bibfield  {author} {\bibinfo {author} {\bibfnamefont {R.}~\bibnamefont
  {Capdessus}}\ and\ \bibinfo {author} {\bibfnamefont {P.}~\bibnamefont
  {McKenna}},\ }\href@noop {} {\bibfield  {journal} {\bibinfo  {journal} {Phys.
  Rev. E}\ }\textbf {\bibinfo {volume} {91}},\ \bibinfo {pages} {053105}
  (\bibinfo {year} {2015})}\BibitemShut {NoStop}%
\bibitem [{\citenamefont {Vranic}\ \emph {et~al.}(2016)\citenamefont {Vranic},
  \citenamefont {Grismayer}, \citenamefont {Fonseca},\ and\ \citenamefont
  {Silva}}]{Vranic_2016}%
  \BibitemOpen
  \bibfield  {author} {\bibinfo {author} {\bibfnamefont {M.}~\bibnamefont
  {Vranic}}, \bibinfo {author} {\bibfnamefont {T.}~\bibnamefont {Grismayer}},
  \bibinfo {author} {\bibfnamefont {R.~A.}\ \bibnamefont {Fonseca}},\ and\
  \bibinfo {author} {\bibfnamefont {L.~O.}\ \bibnamefont {Silva}},\ }\href@noop
  {} {\bibfield  {journal} {\bibinfo  {journal} {New J. Phys.}\ }\textbf
  {\bibinfo {volume} {18}},\ \bibinfo {pages} {073035} (\bibinfo {year}
  {2016})}\BibitemShut {NoStop}%
\bibitem [{\citenamefont {Dinu}\ \emph {et~al.}(2016)\citenamefont {Dinu},
  \citenamefont {Harvey}, \citenamefont {Ilderton}, \citenamefont {Marklund},\
  and\ \citenamefont {Torgrimsson}}]{Dinu_2016}%
  \BibitemOpen
  \bibfield  {author} {\bibinfo {author} {\bibfnamefont {V.}~\bibnamefont
  {Dinu}}, \bibinfo {author} {\bibfnamefont {C.}~\bibnamefont {Harvey}},
  \bibinfo {author} {\bibfnamefont {A.}~\bibnamefont {Ilderton}}, \bibinfo
  {author} {\bibfnamefont {M.}~\bibnamefont {Marklund}},\ and\ \bibinfo
  {author} {\bibfnamefont {G.}~\bibnamefont {Torgrimsson}},\ }\href@noop {}
  {\bibfield  {journal} {\bibinfo  {journal} {Phys. Rev. Lett.}\ }\textbf
  {\bibinfo {volume} {116}},\ \bibinfo {pages} {044801} (\bibinfo {year}
  {2016})}\BibitemShut {NoStop}%
\bibitem [{\citenamefont {Di~Piazza}\ \emph {et~al.}(2017)\citenamefont
  {Di~Piazza}, \citenamefont {Wistisen},\ and\ \citenamefont
  {Uggerh{\o}j}}]{Di_Piazza_2017}%
  \BibitemOpen
  \bibfield  {author} {\bibinfo {author} {\bibfnamefont {A.}~\bibnamefont
  {Di~Piazza}}, \bibinfo {author} {\bibfnamefont {T.~N.}\ \bibnamefont
  {Wistisen}},\ and\ \bibinfo {author} {\bibfnamefont {U.~I.}\ \bibnamefont
  {Uggerh{\o}j}},\ }\href@noop {} {\bibfield  {journal} {\bibinfo  {journal}
  {Phys. Lett. B}\ }\textbf {\bibinfo {volume} {765}},\ \bibinfo {pages} {1}
  (\bibinfo {year} {2017})}\BibitemShut {NoStop}%
\bibitem [{\citenamefont {Harvey}\ \emph {et~al.}(2017)\citenamefont {Harvey},
  \citenamefont {Gonoskov}, \citenamefont {Ilderton},\ and\ \citenamefont
  {Marklund}}]{Harvey_2017}%
  \BibitemOpen
  \bibfield  {author} {\bibinfo {author} {\bibfnamefont {C.~N.}\ \bibnamefont
  {Harvey}}, \bibinfo {author} {\bibfnamefont {A.}~\bibnamefont {Gonoskov}},
  \bibinfo {author} {\bibfnamefont {A.}~\bibnamefont {Ilderton}},\ and\
  \bibinfo {author} {\bibfnamefont {M.}~\bibnamefont {Marklund}},\ }\href@noop
  {} {\bibfield  {journal} {\bibinfo  {journal} {Phys. Rev. Lett.}\ }\textbf
  {\bibinfo {volume} {118}},\ \bibinfo {pages} {105004} (\bibinfo {year}
  {2017})}\BibitemShut {NoStop}%
\bibitem [{\citenamefont {Ridgers}\ \emph {et~al.}(2017)\citenamefont
  {Ridgers}, \citenamefont {Blackburn}, \citenamefont {Del~Sorbo},
  \citenamefont {Bradley}, \citenamefont {Slade-Lowther}, \citenamefont
  {Baird}, \citenamefont {Mangles}, \citenamefont {McKenna}, \citenamefont
  {Marklund}, \citenamefont {Murphy},\ and\ \citenamefont
  {Thomas}}]{Ridgers_2017}%
  \BibitemOpen
  \bibfield  {author} {\bibinfo {author} {\bibfnamefont {C.~P.}\ \bibnamefont
  {Ridgers}}, \bibinfo {author} {\bibfnamefont {T.~G.}\ \bibnamefont
  {Blackburn}}, \bibinfo {author} {\bibfnamefont {D.}~\bibnamefont
  {Del~Sorbo}}, \bibinfo {author} {\bibfnamefont {L.~E.}\ \bibnamefont
  {Bradley}}, \bibinfo {author} {\bibfnamefont {C.}~\bibnamefont
  {Slade-Lowther}}, \bibinfo {author} {\bibfnamefont {C.~D.}\ \bibnamefont
  {Baird}}, \bibinfo {author} {\bibfnamefont {S.~P.~D.}\ \bibnamefont
  {Mangles}}, \bibinfo {author} {\bibfnamefont {P.}~\bibnamefont {McKenna}},
  \bibinfo {author} {\bibfnamefont {M.}~\bibnamefont {Marklund}}, \bibinfo
  {author} {\bibfnamefont {C.~D.}\ \bibnamefont {Murphy}},\ and\ \bibinfo
  {author} {\bibfnamefont {A.~G.~R.}\ \bibnamefont {Thomas}},\ }\href@noop {}
  {\bibfield  {journal} {\bibinfo  {journal} {J. Plasma Phys.}\ }\textbf
  {\bibinfo {volume} {83}},\ \bibinfo {pages} {715830502} (\bibinfo {year}
  {2017})}\BibitemShut {NoStop}%
\bibitem [{\citenamefont {Niel}\ \emph
  {et~al.}(2018{\natexlab{a}})\citenamefont {Niel}, \citenamefont {Riconda},
  \citenamefont {Amiranoff}, \citenamefont {Duclous},\ and\ \citenamefont
  {Grech}}]{Niel_2018a}%
  \BibitemOpen
  \bibfield  {author} {\bibinfo {author} {\bibfnamefont {F.}~\bibnamefont
  {Niel}}, \bibinfo {author} {\bibfnamefont {C.}~\bibnamefont {Riconda}},
  \bibinfo {author} {\bibfnamefont {F.}~\bibnamefont {Amiranoff}}, \bibinfo
  {author} {\bibfnamefont {R.}~\bibnamefont {Duclous}},\ and\ \bibinfo {author}
  {\bibfnamefont {M.}~\bibnamefont {Grech}},\ }\href@noop {} {\bibfield
  {journal} {\bibinfo  {journal} {Phys. Rev. E}\ }\textbf {\bibinfo {volume}
  {97}},\ \bibinfo {pages} {043209} (\bibinfo {year}
  {2018}{\natexlab{a}})}\BibitemShut {NoStop}%
\bibitem [{\citenamefont {Niel}\ \emph
  {et~al.}(2018{\natexlab{b}})\citenamefont {Niel}, \citenamefont {Riconda},
  \citenamefont {Amiranoff}, \citenamefont {Lobet}, \citenamefont {Derouillat},
  \citenamefont {P\'{e}rez}, \citenamefont {Vinci},\ and\ \citenamefont
  {Grech}}]{Niel_2018b}%
  \BibitemOpen
  \bibfield  {author} {\bibinfo {author} {\bibfnamefont {F.}~\bibnamefont
  {Niel}}, \bibinfo {author} {\bibfnamefont {C.}~\bibnamefont {Riconda}},
  \bibinfo {author} {\bibfnamefont {F.}~\bibnamefont {Amiranoff}}, \bibinfo
  {author} {\bibfnamefont {M.}~\bibnamefont {Lobet}}, \bibinfo {author}
  {\bibfnamefont {J.}~\bibnamefont {Derouillat}}, \bibinfo {author}
  {\bibfnamefont {F.}~\bibnamefont {P\'{e}rez}}, \bibinfo {author}
  {\bibfnamefont {T.}~\bibnamefont {Vinci}},\ and\ \bibinfo {author}
  {\bibfnamefont {M.}~\bibnamefont {Grech}},\ }\href@noop {} {\bibfield
  {journal} {\bibinfo  {journal} {Plasma Phys. Controlled Fusion}\ }\textbf
  {\bibinfo {volume} {60}},\ \bibinfo {pages} {094002} (\bibinfo {year}
  {2018}{\natexlab{b}})}\BibitemShut {NoStop}%
\bibitem [{\citenamefont {Wistisen}\ \emph {et~al.}(2018)\citenamefont
  {Wistisen}, \citenamefont {Di~Piazza}, \citenamefont {Knudsen},\ and\
  \citenamefont {Uggerh{\o}j}}]{Wistisen_2018}%
  \BibitemOpen
  \bibfield  {author} {\bibinfo {author} {\bibfnamefont {T.~N.}\ \bibnamefont
  {Wistisen}}, \bibinfo {author} {\bibfnamefont {A.}~\bibnamefont {Di~Piazza}},
  \bibinfo {author} {\bibfnamefont {H.~V.}\ \bibnamefont {Knudsen}},\ and\
  \bibinfo {author} {\bibfnamefont {U.~I.}\ \bibnamefont {Uggerh{\o}j}},\
  }\href@noop {} {\bibfield  {journal} {\bibinfo  {journal} {Nat. Commun.}\
  }\textbf {\bibinfo {volume} {9}},\ \bibinfo {pages} {795} (\bibinfo {year}
  {2018})}\BibitemShut {NoStop}%
\bibitem [{\citenamefont {Cole}\ \emph {et~al.}(2018)\citenamefont {Cole},
  \citenamefont {Behm}, \citenamefont {Gerstmayr}, \citenamefont {Blackburn},
  \citenamefont {Wood}, \citenamefont {Baird}, \citenamefont {Duff},
  \citenamefont {Harvey}, \citenamefont {Ilderton}, \citenamefont {Joglekar},
  \citenamefont {Krushelnick}, \citenamefont {Kuschel}, \citenamefont
  {Marklund}, \citenamefont {McKenna}, \citenamefont {Murphy}, \citenamefont
  {Poder}, \citenamefont {Ridgers}, \citenamefont {Samarin}, \citenamefont
  {Sarri}, \citenamefont {Symes}, \citenamefont {Thomas}, \citenamefont
  {Warwick}, \citenamefont {Zepf}, \citenamefont {Najmudin},\ and\
  \citenamefont {Mangles}}]{Cole_2018}%
  \BibitemOpen
  \bibfield  {author} {\bibinfo {author} {\bibfnamefont {J.~M.}\ \bibnamefont
  {Cole}}, \bibinfo {author} {\bibfnamefont {K.~T.}\ \bibnamefont {Behm}},
  \bibinfo {author} {\bibfnamefont {E.}~\bibnamefont {Gerstmayr}}, \bibinfo
  {author} {\bibfnamefont {T.~G.}\ \bibnamefont {Blackburn}}, \bibinfo {author}
  {\bibfnamefont {J.~C.}\ \bibnamefont {Wood}}, \bibinfo {author}
  {\bibfnamefont {C.~D.}\ \bibnamefont {Baird}}, \bibinfo {author}
  {\bibfnamefont {M.~J.}\ \bibnamefont {Duff}}, \bibinfo {author}
  {\bibfnamefont {C.}~\bibnamefont {Harvey}}, \bibinfo {author} {\bibfnamefont
  {A.}~\bibnamefont {Ilderton}}, \bibinfo {author} {\bibfnamefont {A.~S.}\
  \bibnamefont {Joglekar}}, \bibinfo {author} {\bibfnamefont {K.}~\bibnamefont
  {Krushelnick}}, \bibinfo {author} {\bibfnamefont {S.}~\bibnamefont
  {Kuschel}}, \bibinfo {author} {\bibfnamefont {M.}~\bibnamefont {Marklund}},
  \bibinfo {author} {\bibfnamefont {P.}~\bibnamefont {McKenna}}, \bibinfo
  {author} {\bibfnamefont {C.~D.}\ \bibnamefont {Murphy}}, \bibinfo {author}
  {\bibfnamefont {K.}~\bibnamefont {Poder}}, \bibinfo {author} {\bibfnamefont
  {C.~P.}\ \bibnamefont {Ridgers}}, \bibinfo {author} {\bibfnamefont {G.~M.}\
  \bibnamefont {Samarin}}, \bibinfo {author} {\bibfnamefont {G.}~\bibnamefont
  {Sarri}}, \bibinfo {author} {\bibfnamefont {D.~R.}\ \bibnamefont {Symes}},
  \bibinfo {author} {\bibfnamefont {A.~G.~R.}\ \bibnamefont {Thomas}}, \bibinfo
  {author} {\bibfnamefont {J.}~\bibnamefont {Warwick}}, \bibinfo {author}
  {\bibfnamefont {M.}~\bibnamefont {Zepf}}, \bibinfo {author} {\bibfnamefont
  {Z.}~\bibnamefont {Najmudin}},\ and\ \bibinfo {author} {\bibfnamefont
  {S.~P.~D.}\ \bibnamefont {Mangles}},\ }\href@noop {} {\bibfield  {journal}
  {\bibinfo  {journal} {Phys. Rev. X}\ }\textbf {\bibinfo {volume} {8}},\
  \bibinfo {pages} {011020} (\bibinfo {year} {2018})}\BibitemShut {NoStop}%
\bibitem [{\citenamefont {Poder}\ \emph {et~al.}(2018)\citenamefont {Poder},
  \citenamefont {Tamburini}, \citenamefont {Sarri}, \citenamefont {Di~Piazza},
  \citenamefont {Kuschel}, \citenamefont {Baird}, \citenamefont {Behm},
  \citenamefont {Bohlen}, \citenamefont {Cole}, \citenamefont {Corvan},
  \citenamefont {Duff}, \citenamefont {Gerstmayr}, \citenamefont {Keitel},
  \citenamefont {Krushelnick}, \citenamefont {Mangles}, \citenamefont
  {McKenna}, \citenamefont {Murphy}, \citenamefont {Najmudin}, \citenamefont
  {Ridgers}, \citenamefont {Samarin}, \citenamefont {Symes}, \citenamefont
  {Thomas}, \citenamefont {Warwick},\ and\ \citenamefont {Zepf}}]{Poder_2018}%
  \BibitemOpen
  \bibfield  {author} {\bibinfo {author} {\bibfnamefont {K.}~\bibnamefont
  {Poder}}, \bibinfo {author} {\bibfnamefont {M.}~\bibnamefont {Tamburini}},
  \bibinfo {author} {\bibfnamefont {G.}~\bibnamefont {Sarri}}, \bibinfo
  {author} {\bibfnamefont {A.}~\bibnamefont {Di~Piazza}}, \bibinfo {author}
  {\bibfnamefont {S.}~\bibnamefont {Kuschel}}, \bibinfo {author} {\bibfnamefont
  {C.~D.}\ \bibnamefont {Baird}}, \bibinfo {author} {\bibfnamefont
  {K.}~\bibnamefont {Behm}}, \bibinfo {author} {\bibfnamefont {S.}~\bibnamefont
  {Bohlen}}, \bibinfo {author} {\bibfnamefont {J.~M.}\ \bibnamefont {Cole}},
  \bibinfo {author} {\bibfnamefont {D.~J.}\ \bibnamefont {Corvan}}, \bibinfo
  {author} {\bibfnamefont {M.}~\bibnamefont {Duff}}, \bibinfo {author}
  {\bibfnamefont {E.}~\bibnamefont {Gerstmayr}}, \bibinfo {author}
  {\bibfnamefont {C.~H.}\ \bibnamefont {Keitel}}, \bibinfo {author}
  {\bibfnamefont {K.}~\bibnamefont {Krushelnick}}, \bibinfo {author}
  {\bibfnamefont {S.~P.~D.}\ \bibnamefont {Mangles}}, \bibinfo {author}
  {\bibfnamefont {P.}~\bibnamefont {McKenna}}, \bibinfo {author} {\bibfnamefont
  {C.~D.}\ \bibnamefont {Murphy}}, \bibinfo {author} {\bibfnamefont
  {Z.}~\bibnamefont {Najmudin}}, \bibinfo {author} {\bibfnamefont {C.~P.}\
  \bibnamefont {Ridgers}}, \bibinfo {author} {\bibfnamefont {G.~M.}\
  \bibnamefont {Samarin}}, \bibinfo {author} {\bibfnamefont {D.~R.}\
  \bibnamefont {Symes}}, \bibinfo {author} {\bibfnamefont {A.~G.~R.}\
  \bibnamefont {Thomas}}, \bibinfo {author} {\bibfnamefont {J.}~\bibnamefont
  {Warwick}},\ and\ \bibinfo {author} {\bibfnamefont {M.}~\bibnamefont
  {Zepf}},\ }\href@noop {} {\bibfield  {journal} {\bibinfo  {journal} {Phys.
  Rev. X}\ }\textbf {\bibinfo {volume} {8}},\ \bibinfo {pages} {031004}
  (\bibinfo {year} {2018})}\BibitemShut {NoStop}%
\bibitem [{\citenamefont {Hammond}(2010)}]{Hammond_2010}%
  \BibitemOpen
  \bibfield  {author} {\bibinfo {author} {\bibfnamefont {R.~T.}\ \bibnamefont
  {Hammond}},\ }\href@noop {} {\bibfield  {journal} {\bibinfo  {journal}
  {Electron. J. Theor. Phys.}\ }\textbf {\bibinfo {volume} {7}},\ \bibinfo
  {pages} {221} (\bibinfo {year} {2010})}\BibitemShut {NoStop}%
\bibitem [{\citenamefont {Di~Piazza}\ \emph {et~al.}(2012)\citenamefont
  {Di~Piazza}, \citenamefont {M\"{u}ller}, \citenamefont {Hatsagortsyan},\ and\
  \citenamefont {Keitel}}]{Di_Piazza_2012}%
  \BibitemOpen
  \bibfield  {author} {\bibinfo {author} {\bibfnamefont {A.}~\bibnamefont
  {Di~Piazza}}, \bibinfo {author} {\bibfnamefont {C.}~\bibnamefont
  {M\"{u}ller}}, \bibinfo {author} {\bibfnamefont {K.~Z.}\ \bibnamefont
  {Hatsagortsyan}},\ and\ \bibinfo {author} {\bibfnamefont {C.~H.}\
  \bibnamefont {Keitel}},\ }\href@noop {} {\bibfield  {journal} {\bibinfo
  {journal} {Rev. Mod. Phys.}\ }\textbf {\bibinfo {volume} {84}},\ \bibinfo
  {pages} {1177} (\bibinfo {year} {2012})}\BibitemShut {NoStop}%
\bibitem [{\citenamefont {Burton}\ and\ \citenamefont
  {Noble}(2014)}]{Burton_2014}%
  \BibitemOpen
  \bibfield  {author} {\bibinfo {author} {\bibfnamefont {D.~A.}\ \bibnamefont
  {Burton}}\ and\ \bibinfo {author} {\bibfnamefont {A.}~\bibnamefont {Noble}},\
  }\href@noop {} {\bibfield  {journal} {\bibinfo  {journal} {Contemp. Phys.}\
  }\textbf {\bibinfo {volume} {55}},\ \bibinfo {pages} {110} (\bibinfo {year}
  {2014})}\BibitemShut {NoStop}%
\bibitem [{\citenamefont {Blackburn}(2020)}]{Blackburn_2020}%
  \BibitemOpen
  \bibfield  {author} {\bibinfo {author} {\bibfnamefont {T.~G.}\ \bibnamefont
  {Blackburn}},\ }\href@noop {} {\bibfield  {journal} {\bibinfo  {journal}
  {Rev. Mod. Plasma Phys.}\ }\textbf {\bibinfo {volume} {4}},\ \bibinfo {pages}
  {5} (\bibinfo {year} {2020})}\BibitemShut {NoStop}%
\bibitem [{\citenamefont {Di~Piazza}(2008)}]{Di_Piazza_2008_a}%
  \BibitemOpen
  \bibfield  {author} {\bibinfo {author} {\bibfnamefont {A.}~\bibnamefont
  {Di~Piazza}},\ }\href@noop {} {\bibfield  {journal} {\bibinfo  {journal}
  {Lett. Math. Phys.}\ }\textbf {\bibinfo {volume} {83}},\ \bibinfo {pages}
  {305} (\bibinfo {year} {2008})}\BibitemShut {NoStop}%
\bibitem [{\citenamefont {Ritus}(1985)}]{Ritus_1985}%
  \BibitemOpen
  \bibfield  {author} {\bibinfo {author} {\bibfnamefont {V.~I.}\ \bibnamefont
  {Ritus}},\ }\href@noop {} {\bibfield  {journal} {\bibinfo  {journal} {J. Sov.
  Laser Res.}\ }\textbf {\bibinfo {volume} {6}},\ \bibinfo {pages} {497}
  (\bibinfo {year} {1985})}\BibitemShut {NoStop}%
\bibitem [{\citenamefont {Baier}\ \emph {et~al.}(1998)\citenamefont {Baier},
  \citenamefont {Katkov},\ and\ \citenamefont {Strakhovenko}}]{Baier_b_1998}%
  \BibitemOpen
  \bibfield  {author} {\bibinfo {author} {\bibfnamefont {V.~N.}\ \bibnamefont
  {Baier}}, \bibinfo {author} {\bibfnamefont {V.~M.}\ \bibnamefont {Katkov}},\
  and\ \bibinfo {author} {\bibfnamefont {V.~M.}\ \bibnamefont {Strakhovenko}},\
  }\href@noop {} {\emph {\bibinfo {title} {Electromagnetic Processes at High
  Energies in Oriented Single Crystals}}}\ (\bibinfo  {publisher} {World
  Scientific, Singapore},\ \bibinfo {year} {1998})\BibitemShut {NoStop}%
\bibitem [{\citenamefont {Di~Piazza}(2018{\natexlab{a}})}]{Di_Piazza_2018_b}%
  \BibitemOpen
  \bibfield  {author} {\bibinfo {author} {\bibfnamefont {A.}~\bibnamefont
  {Di~Piazza}},\ }\href@noop {} {\bibfield  {journal} {\bibinfo  {journal}
  {Phys. Lett. B}\ }\textbf {\bibinfo {volume} {782}},\ \bibinfo {pages} {559}
  (\bibinfo {year} {2018}{\natexlab{a}})}\BibitemShut {NoStop}%
\bibitem [{\citenamefont {Khokonov}(2019)}]{Khokonov_2019}%
  \BibitemOpen
  \bibfield  {author} {\bibinfo {author} {\bibfnamefont {M.~K.}\ \bibnamefont
  {Khokonov}},\ }\href@noop {} {\bibfield  {journal} {\bibinfo  {journal}
  {Phys. Lett. B}\ }\textbf {\bibinfo {volume} {791}},\ \bibinfo {pages} {281}
  (\bibinfo {year} {2019})}\BibitemShut {NoStop}%
\bibitem [{\citenamefont {Di~Piazza}\ \emph {et~al.}(2018)\citenamefont
  {Di~Piazza}, \citenamefont {Tamburini}, \citenamefont {Meuren},\ and\
  \citenamefont {Keitel}}]{Di_Piazza_2018}%
  \BibitemOpen
  \bibfield  {author} {\bibinfo {author} {\bibfnamefont {A.}~\bibnamefont
  {Di~Piazza}}, \bibinfo {author} {\bibfnamefont {M.}~\bibnamefont
  {Tamburini}}, \bibinfo {author} {\bibfnamefont {S.}~\bibnamefont {Meuren}},\
  and\ \bibinfo {author} {\bibfnamefont {C.~H.}\ \bibnamefont {Keitel}},\
  }\href@noop {} {\bibfield  {journal} {\bibinfo  {journal} {Phys. Rev. A}\
  }\textbf {\bibinfo {volume} {98}},\ \bibinfo {pages} {012134} (\bibinfo
  {year} {2018})}\BibitemShut {NoStop}%
\bibitem [{\citenamefont {Glauber}(1951)}]{Glauber_1951}%
  \BibitemOpen
  \bibfield  {author} {\bibinfo {author} {\bibfnamefont {R.~J.}\ \bibnamefont
  {Glauber}},\ }\href@noop {} {\bibfield  {journal} {\bibinfo  {journal} {Phys.
  Rev.}\ }\textbf {\bibinfo {volume} {84}},\ \bibinfo {pages} {395} (\bibinfo
  {year} {1951})}\BibitemShut {NoStop}%
\bibitem [{\citenamefont {Di~Piazza}\ \emph {et~al.}(2010)\citenamefont
  {Di~Piazza}, \citenamefont {Hatsagortsyan},\ and\ \citenamefont
  {Keitel}}]{Di_Piazza_2010}%
  \BibitemOpen
  \bibfield  {author} {\bibinfo {author} {\bibfnamefont {A.}~\bibnamefont
  {Di~Piazza}}, \bibinfo {author} {\bibfnamefont {K.~Z.}\ \bibnamefont
  {Hatsagortsyan}},\ and\ \bibinfo {author} {\bibfnamefont {C.~H.}\
  \bibnamefont {Keitel}},\ }\href@noop {} {\bibfield  {journal} {\bibinfo
  {journal} {Phys. Rev. Lett.}\ }\textbf {\bibinfo {volume} {105}},\ \bibinfo
  {pages} {220403} (\bibinfo {year} {2010})}\BibitemShut {NoStop}%
\bibitem [{\citenamefont {Tamburini}\ and\ \citenamefont
  {Meuren}()}]{Tamburini_2019}%
  \BibitemOpen
  \bibfield  {author} {\bibinfo {author} {\bibfnamefont {M.}~\bibnamefont
  {Tamburini}}\ and\ \bibinfo {author} {\bibfnamefont {S.}~\bibnamefont
  {Meuren}},\ }\href@noop {} {\bibinfo  {journal} {arXiv:1912.07508}\
  }\BibitemShut {NoStop}%
\bibitem [{\citenamefont {Brown}\ and\ \citenamefont
  {Kibble}(1964)}]{Brown_1964}%
  \BibitemOpen
\bibfield  {journal} {  }\bibfield  {author} {\bibinfo {author} {\bibfnamefont
  {L.~S.}\ \bibnamefont {Brown}}\ and\ \bibinfo {author} {\bibfnamefont
  {T.~W.~B.}\ \bibnamefont {Kibble}},\ }\href@noop {} {\bibfield  {journal}
  {\bibinfo  {journal} {Phys. Rev.}\ }\textbf {\bibinfo {volume} {133}},\
  \bibinfo {pages} {A705} (\bibinfo {year} {1964})}\BibitemShut {NoStop}%
\bibitem [{\citenamefont {Kibble}\ \emph {et~al.}(1975)\citenamefont {Kibble},
  \citenamefont {Salam},\ and\ \citenamefont {Strathdee}}]{Kibble_1975}%
  \BibitemOpen
  \bibfield  {author} {\bibinfo {author} {\bibfnamefont {T.~W.~B.}\
  \bibnamefont {Kibble}}, \bibinfo {author} {\bibfnamefont {A.}~\bibnamefont
  {Salam}},\ and\ \bibinfo {author} {\bibfnamefont {J.}~\bibnamefont
  {Strathdee}},\ }\href@noop {} {\bibfield  {journal} {\bibinfo  {journal}
  {Nucl. Phys.}\ }\textbf {\bibinfo {volume} {B96}},\ \bibinfo {pages} {255}
  (\bibinfo {year} {1975})}\BibitemShut {NoStop}%
\bibitem [{\citenamefont {Hebenstreit}\ \emph {et~al.}(2011)\citenamefont
  {Hebenstreit}, \citenamefont {Ilderton}, \citenamefont {Marklund},\ and\
  \citenamefont {Zamanian}}]{Hebenstreit_2011}%
  \BibitemOpen
  \bibfield  {author} {\bibinfo {author} {\bibfnamefont {F.}~\bibnamefont
  {Hebenstreit}}, \bibinfo {author} {\bibfnamefont {A.}~\bibnamefont
  {Ilderton}}, \bibinfo {author} {\bibfnamefont {M.}~\bibnamefont {Marklund}},\
  and\ \bibinfo {author} {\bibfnamefont {J.}~\bibnamefont {Zamanian}},\
  }\href@noop {} {\bibfield  {journal} {\bibinfo  {journal} {Phys. Rev. D}\
  }\textbf {\bibinfo {volume} {83}},\ \bibinfo {pages} {065007} (\bibinfo
  {year} {2011})}\BibitemShut {NoStop}%
\bibitem [{\citenamefont {Di~Piazza}(2018{\natexlab{b}})}]{Di_Piazza_2018_d}%
  \BibitemOpen
  \bibfield  {author} {\bibinfo {author} {\bibfnamefont {A.}~\bibnamefont
  {Di~Piazza}},\ }\href@noop {} {\bibfield  {journal} {\bibinfo  {journal}
  {Phys. Rev. D}\ }\textbf {\bibinfo {volume} {97}},\ \bibinfo {pages} {056028}
  (\bibinfo {year} {2018}{\natexlab{b}})}\BibitemShut {NoStop}%
\bibitem [{\citenamefont {Baier}\ \emph {et~al.}(1989)\citenamefont {Baier},
  \citenamefont {Katkov},\ and\ \citenamefont {Strakhovenko}}]{Baier_1989}%
  \BibitemOpen
  \bibfield  {author} {\bibinfo {author} {\bibfnamefont {V.~N.}\ \bibnamefont
  {Baier}}, \bibinfo {author} {\bibfnamefont {V.~M.}\ \bibnamefont {Katkov}},\
  and\ \bibinfo {author} {\bibfnamefont {V.~M.}\ \bibnamefont {Strakhovenko}},\
  }\href@noop {} {\bibfield  {journal} {\bibinfo  {journal} {Nucl. Phys.}\
  }\textbf {\bibinfo {volume} {B328}},\ \bibinfo {pages} {387} (\bibinfo {year}
  {1989})}\BibitemShut {NoStop}%
\bibitem [{\citenamefont {Khokonov}\ and\ \citenamefont
  {Nitta}(2002)}]{Khokonov_2002}%
  \BibitemOpen
  \bibfield  {author} {\bibinfo {author} {\bibfnamefont {M.~K.}\ \bibnamefont
  {Khokonov}}\ and\ \bibinfo {author} {\bibfnamefont {H.}~\bibnamefont
  {Nitta}},\ }\href@noop {} {\bibfield  {journal} {\bibinfo  {journal} {Phys.
  Rev. Lett.}\ }\textbf {\bibinfo {volume} {89}},\ \bibinfo {pages} {094801}
  (\bibinfo {year} {2002})}\BibitemShut {NoStop}%
\bibitem [{\citenamefont {Di~Piazza}\ \emph {et~al.}(2007)\citenamefont
  {Di~Piazza}, \citenamefont {Milstein},\ and\ \citenamefont
  {Keitel}}]{Di_Piazza_2007}%
  \BibitemOpen
  \bibfield  {author} {\bibinfo {author} {\bibfnamefont {A.}~\bibnamefont
  {Di~Piazza}}, \bibinfo {author} {\bibfnamefont {A.~I.}\ \bibnamefont
  {Milstein}},\ and\ \bibinfo {author} {\bibfnamefont {C.~H.}\ \bibnamefont
  {Keitel}},\ }\href@noop {} {\bibfield  {journal} {\bibinfo  {journal} {Phys.
  Rev. A}\ }\textbf {\bibinfo {volume} {76}},\ \bibinfo {pages} {032103}
  (\bibinfo {year} {2007})}\BibitemShut {NoStop}%
\bibitem [{\citenamefont {Wistisen}(2015)}]{Wistisen_2015}%
  \BibitemOpen
  \bibfield  {author} {\bibinfo {author} {\bibfnamefont {T.~N.}\ \bibnamefont
  {Wistisen}},\ }\href@noop {} {\bibfield  {journal} {\bibinfo  {journal}
  {Phys. Rev. D}\ }\textbf {\bibinfo {volume} {92}},\ \bibinfo {pages} {045045}
  (\bibinfo {year} {2015})}\BibitemShut {NoStop}%
\bibitem [{\citenamefont {Harvey}\ \emph {et~al.}(2015)\citenamefont {Harvey},
  \citenamefont {Ilderton},\ and\ \citenamefont {King}}]{Harvey_2015}%
  \BibitemOpen
  \bibfield  {author} {\bibinfo {author} {\bibfnamefont {C.~N.}\ \bibnamefont
  {Harvey}}, \bibinfo {author} {\bibfnamefont {A.}~\bibnamefont {Ilderton}},\
  and\ \bibinfo {author} {\bibfnamefont {B.}~\bibnamefont {King}},\ }\href@noop
  {} {\bibfield  {journal} {\bibinfo  {journal} {Phys. Rev. A}\ }\textbf
  {\bibinfo {volume} {91}},\ \bibinfo {pages} {013822} (\bibinfo {year}
  {2015})}\BibitemShut {NoStop}%
\bibitem [{\citenamefont {Blackburn}\ \emph {et~al.}(2018)\citenamefont
  {Blackburn}, \citenamefont {Seipt}, \citenamefont {Bulanov},\ and\
  \citenamefont {Marklund}}]{Blackburn_2018}%
  \BibitemOpen
  \bibfield  {author} {\bibinfo {author} {\bibfnamefont {T.~G.}\ \bibnamefont
  {Blackburn}}, \bibinfo {author} {\bibfnamefont {D.}~\bibnamefont {Seipt}},
  \bibinfo {author} {\bibfnamefont {S.~S.}\ \bibnamefont {Bulanov}},\ and\
  \bibinfo {author} {\bibfnamefont {M.}~\bibnamefont {Marklund}},\ }\href@noop
  {} {\bibfield  {journal} {\bibinfo  {journal} {Phys. Plasmas}\ }\textbf
  {\bibinfo {volume} {25}},\ \bibinfo {pages} {083108} (\bibinfo {year}
  {2018})}\BibitemShut {NoStop}%
\bibitem [{\citenamefont {Aleksandrov}\ \emph {et~al.}(2019)\citenamefont
  {Aleksandrov}, \citenamefont {Plunien},\ and\ \citenamefont
  {Shabaev}}]{Alexandrov_2019}%
  \BibitemOpen
  \bibfield  {author} {\bibinfo {author} {\bibfnamefont {I.~A.}\ \bibnamefont
  {Aleksandrov}}, \bibinfo {author} {\bibfnamefont {G.}~\bibnamefont
  {Plunien}},\ and\ \bibinfo {author} {\bibfnamefont {V.~M.}\ \bibnamefont
  {Shabaev}},\ }\href@noop {} {\bibfield  {journal} {\bibinfo  {journal} {Phys.
  Rev. D}\ }\textbf {\bibinfo {volume} {99}},\ \bibinfo {pages} {016020}
  (\bibinfo {year} {2019})}\BibitemShut {NoStop}%
\bibitem [{\citenamefont {Di~Piazza}\ \emph {et~al.}(2019)\citenamefont
  {Di~Piazza}, \citenamefont {Tamburini}, \citenamefont {Meuren},\ and\
  \citenamefont {Keitel}}]{Di_Piazza_2019}%
  \BibitemOpen
  \bibfield  {author} {\bibinfo {author} {\bibfnamefont {A.}~\bibnamefont
  {Di~Piazza}}, \bibinfo {author} {\bibfnamefont {M.}~\bibnamefont
  {Tamburini}}, \bibinfo {author} {\bibfnamefont {S.}~\bibnamefont {Meuren}},\
  and\ \bibinfo {author} {\bibfnamefont {C.~H.}\ \bibnamefont {Keitel}},\
  }\href@noop {} {\bibfield  {journal} {\bibinfo  {journal} {Phys. Rev. A}\
  }\textbf {\bibinfo {volume} {99}},\ \bibinfo {pages} {022125} (\bibinfo
  {year} {2019})}\BibitemShut {NoStop}%
\bibitem [{\citenamefont {Ilderton}\ \emph {et~al.}(2019)\citenamefont
  {Ilderton}, \citenamefont {King},\ and\ \citenamefont
  {Seipt}}]{Ilderton_2019_b}%
  \BibitemOpen
  \bibfield  {author} {\bibinfo {author} {\bibfnamefont {A.}~\bibnamefont
  {Ilderton}}, \bibinfo {author} {\bibfnamefont {B.}~\bibnamefont {King}},\
  and\ \bibinfo {author} {\bibfnamefont {D.}~\bibnamefont {Seipt}},\
  }\href@noop {} {\bibfield  {journal} {\bibinfo  {journal} {Phys. Rev. A}\
  }\textbf {\bibinfo {volume} {99}},\ \bibinfo {pages} {042121} (\bibinfo
  {year} {2019})}\BibitemShut {NoStop}%
\bibitem [{\citenamefont {Podszus}\ and\ \citenamefont
  {Di~Piazza}(2019)}]{Podszus_2019}%
  \BibitemOpen
  \bibfield  {author} {\bibinfo {author} {\bibfnamefont {T.}~\bibnamefont
  {Podszus}}\ and\ \bibinfo {author} {\bibfnamefont {A.}~\bibnamefont
  {Di~Piazza}},\ }\href@noop {} {\bibfield  {journal} {\bibinfo  {journal}
  {Phys. Rev. D}\ }\textbf {\bibinfo {volume} {99}},\ \bibinfo {pages} {076004}
  (\bibinfo {year} {2019})}\BibitemShut {NoStop}%
\bibitem [{\citenamefont {Ilderton}(2019)}]{Ilderton_2019}%
  \BibitemOpen
  \bibfield  {author} {\bibinfo {author} {\bibfnamefont {A.}~\bibnamefont
  {Ilderton}},\ }\href@noop {} {\bibfield  {journal} {\bibinfo  {journal}
  {Phys. Rev. D}\ }\textbf {\bibinfo {volume} {99}},\ \bibinfo {pages} {085002}
  (\bibinfo {year} {2019})}\BibitemShut {NoStop}%
\bibitem [{\citenamefont {Lv}\ \emph {et~al.}(2021)\citenamefont {Lv},
  \citenamefont {Raicher}, \citenamefont {Keitel},\ and\ \citenamefont
  {Hatsagortsyan}}]{Lv_2021}%
  \BibitemOpen
  \bibfield  {author} {\bibinfo {author} {\bibfnamefont {Q.~Z.}\ \bibnamefont
  {Lv}}, \bibinfo {author} {\bibfnamefont {E.}~\bibnamefont {Raicher}},
  \bibinfo {author} {\bibfnamefont {C.~H.}\ \bibnamefont {Keitel}},\ and\
  \bibinfo {author} {\bibfnamefont {K.~Z.}\ \bibnamefont {Hatsagortsyan}},\
  }\href@noop {} {\bibfield  {journal} {\bibinfo  {journal} {Phys. Rev.
  Research}\ }\textbf {\bibinfo {volume} {3}},\ \bibinfo {pages} {013214}
  (\bibinfo {year} {2021})}\BibitemShut {NoStop}%
\bibitem [{\citenamefont {Olver}\ \emph {et~al.}(2010)\citenamefont {Olver},
  \citenamefont {Lozier}, \citenamefont {Boisvert},\ and\ \citenamefont
  {Clark}}]{NIST_b_2010}%
  \BibitemOpen
  \bibinfo {editor} {\bibfnamefont {F.~W.~J.}\ \bibnamefont {Olver}}, \bibinfo
  {editor} {\bibfnamefont {D.~W.}\ \bibnamefont {Lozier}}, \bibinfo {editor}
  {\bibfnamefont {R.~F.}\ \bibnamefont {Boisvert}},\ and\ \bibinfo {editor}
  {\bibfnamefont {C.~W.}\ \bibnamefont {Clark}},\ eds.,\ \href@noop {} {\emph
  {\bibinfo {title} {NIST Handbook of Mathematical Functions}}}\ (\bibinfo
  {publisher} {Cambridge University Press, Cambridge, England},\ \bibinfo
  {year} {2010})\BibitemShut {NoStop}%
\bibitem [{\citenamefont {Heinzl}\ \emph {et~al.}()\citenamefont {Heinzl},
  \citenamefont {Ilderton},\ and\ \citenamefont {King}}]{Heinzl_2021}%
  \BibitemOpen
  \bibfield  {author} {\bibinfo {author} {\bibfnamefont {T.}~\bibnamefont
  {Heinzl}}, \bibinfo {author} {\bibfnamefont {A.}~\bibnamefont {Ilderton}},\
  and\ \bibinfo {author} {\bibfnamefont {B.}~\bibnamefont {King}},\ }\href@noop
  {} {\bibinfo  {journal} {arXiv:2101.12111}\ }\BibitemShut {NoStop}%
\bibitem [{\citenamefont {Ritus}(1970)}]{Ritus_1970}%
  \BibitemOpen
\bibfield  {journal} {  }\bibfield  {author} {\bibinfo {author} {\bibfnamefont
  {V.~I.}\ \bibnamefont {Ritus}},\ }\href@noop {} {\bibfield  {journal}
  {\bibinfo  {journal} {Sov. Phys. JETP}\ }\textbf {\bibinfo {volume} {30}},\
  \bibinfo {pages} {1181} (\bibinfo {year} {1970})}\BibitemShut {NoStop}%
\bibitem [{\citenamefont {Narozhny}(1979)}]{Narozhny_1979}%
  \BibitemOpen
  \bibfield  {author} {\bibinfo {author} {\bibfnamefont {N.~B.}\ \bibnamefont
  {Narozhny}},\ }\href@noop {} {\bibfield  {journal} {\bibinfo  {journal}
  {Phys. Rev. D}\ }\textbf {\bibinfo {volume} {20}},\ \bibinfo {pages} {1313}
  (\bibinfo {year} {1979})}\BibitemShut {NoStop}%
\bibitem [{\citenamefont {Narozhny}(1980)}]{Narozhny_1980}%
  \BibitemOpen
  \bibfield  {author} {\bibinfo {author} {\bibfnamefont {N.~B.}\ \bibnamefont
  {Narozhny}},\ }\href@noop {} {\bibfield  {journal} {\bibinfo  {journal}
  {Phys. Rev. D}\ }\textbf {\bibinfo {volume} {21}},\ \bibinfo {pages} {1176}
  (\bibinfo {year} {1980})}\BibitemShut {NoStop}%
\bibitem [{\citenamefont {Morozov}\ \emph {et~al.}(1981)\citenamefont
  {Morozov}, \citenamefont {Narozhny},\ and\ \citenamefont
  {Ritus}}]{Morozov_1981}%
  \BibitemOpen
  \bibfield  {author} {\bibinfo {author} {\bibfnamefont {D.~A.}\ \bibnamefont
  {Morozov}}, \bibinfo {author} {\bibfnamefont {N.~B.}\ \bibnamefont
  {Narozhny}},\ and\ \bibinfo {author} {\bibfnamefont {V.~I.}\ \bibnamefont
  {Ritus}},\ }\href@noop {} {\bibfield  {journal} {\bibinfo  {journal} {Sov.
  Phys. JETP}\ }\textbf {\bibinfo {volume} {53}},\ \bibinfo {pages} {1103}
  (\bibinfo {year} {1981})}\BibitemShut {NoStop}%
\bibitem [{\citenamefont {Akhmedov}(2011)}]{Akhmedov_2011}%
  \BibitemOpen
  \bibfield  {author} {\bibinfo {author} {\bibfnamefont {E.~K.}\ \bibnamefont
  {Akhmedov}},\ }\href@noop {} {\bibfield  {journal} {\bibinfo  {journal}
  {Phys. At. Nucl.}\ }\textbf {\bibinfo {volume} {74}},\ \bibinfo {pages}
  {1299} (\bibinfo {year} {2011})}\BibitemShut {NoStop}%
\bibitem [{\citenamefont {Fedotov}(2017)}]{Fedotov_2017}%
  \BibitemOpen
  \bibfield  {author} {\bibinfo {author} {\bibfnamefont {A.~M.}\ \bibnamefont
  {Fedotov}},\ }\href@noop {} {\bibfield  {journal} {\bibinfo  {journal} {J.
  Phys. Conf. Ser.}\ }\textbf {\bibinfo {volume} {826}},\ \bibinfo {pages}
  {012027} (\bibinfo {year} {2017})}\BibitemShut {NoStop}%
\end{thebibliography}

\end{document}